\documentclass[12pt]{article}

\newcommand{\Tr}{{\rm Tr}\,}
\newcommand{\sub}[1]{_{\rm #1}}
\renewcommand{\top}[1]{^{\rm #1}}
\newcommand{\pkt}{\;.}
\newcommand{\kma}{\;,}
\newcommand{\tr}{\,{\rm tr}\,}
\newcommand{\bea}{\begin{eqnarray}}
\newcommand{\be}{\begin{equation}}
\newcommand{\eea}{\end{eqnarray}}
\newcommand{\ee}{\end{equation}}
\newcommand{\btau}{\mbox{\boldmath{$\tau$}\unboldmath}}
\newcommand{\bOmega}{\mbox{\boldmath{$\Omega$}\unboldmath}}

\newcommand{\bpi}{\mbox{\boldmath{$\pi$}\unboldmath}}
\newcommand{\bphi}{\mbox{\boldmath{$\phi$}\unboldmath}}

\newcommand{\hbphi}{\mbox{\boldmath{$\hat{\phi}$}\unboldmath}}
\newcommand{\bx}{\mbox{\boldmath{$x$}\unboldmath}}

\newcommand{\bM}{\mbox{\boldmath{$M$}\unboldmath}}
\newcommand{\bgamma}{\mbox{\boldmath{$\gamma$}\unboldmath}}
\newcommand{\balpha}{\mbox{\boldmath{$\alpha$}\unboldmath}}  

\newcommand{\bnabla}{\mbox{\boldmath{$\nabla$}\unboldmath}}

\newcommand{\bz}{\mbox{\boldmath{$z$}\unboldmath}}

\newcommand{\bOmegaA}{\mbox{\boldmath{$\scriptstyle \Omega$}\unboldmath}}      
\newcommand{\bq}{\mbox{\boldmath{$q$}\unboldmath}}

\newcommand{\hbz}{\mbox{\boldmath{$\hat z$}\unboldmath}}

\newcommand{\hbx}{\mbox{\boldmath{$\hat x$}\unboldmath}}

\newcommand{\lag}{{\cal L}}
\begin{document}
\begin{titlepage}
\begin{flushright}
DO-TH-98/19\\
\end{flushright}

\vspace{18mm}
\begin{center}
{\Large \bf
The Nucleon in the chiral quark model:\\
 an alternative 
computation scheme}

\vspace{8mm}
\renewcommand{\thefootnote}{\fnsymbol{footnote}}

{\large  
J. Baacke\footnote{{Email address:~baacke@physik.uni-dortmund.de}}
and H. Sprenger\footnote{Email address:
sprenger@hal1.physik.uni-dortmund.de}
}\\
\vspace{10mm}

{\large Institut f\"ur Physik, Universit\"at Dortmund} \\
{\large D - 44221 Dortmund, Germany}
\vspace{15mm}
\end{center}
\begin{center}
\parbox[t]{14cm}
{\begin{center}
{\bf{Abstract}}
\end{center}
We formulate the quark sea contributions to the energy and
various physical observables in terms of Euclidean Green functions
and their K-spin partial wave reduction. In this framework
it is not necessary to discretize the continuous spectrum
by introducing a finite boundary. 
Using this formulation we perform a new self-consistent 
computation of the nucleon state in the Nambu-Jona-Lasinio 
model. Besides this technical advantage such an alternative
 computation scheme makes it possible to obtain 
the numerical predictions of the model in an entirely independent way. 
We use the Pauli-Villars cutoff to define the model.
Our results for the nucleon energy, the mesonic profile, and 
various observables essentially confirm results
obtained by other groups using this regularization. The 
results for $g\sub{A}$, obtained on the basis of quark currents,
are close to the experimental value.}
\end{center}
\vspace{.8cm}

\noindent
{\it Keywords:}

\noindent
non-perturbative methods, effective chiral models, 
chiral solitons, models of baryons

\noindent
{\it PACS:} 

\noindent
12.38.Lg, 12.39.Fe, 12.40.Yx
\end{titlepage}

\section{Introduction}
\setcounter{equation}{0}

The Nambu-Jona-Lasinio (NJL) model  \cite{Nambu:1961a}
has received a wide attention during
the last decade \cite{Alkofer:1996},
following previous attempts 
\cite{Skyrme:1961,Adkins:1983} to describe the
nucleon in the large-$N\sub{c}$ limit on the basis of mesonic 
degrees of freedom. In the NJL model, the mesons appear as
effective degrees of freedom, parametrizing condensates of
the basic fermion fields. The basic model is a
quark model with a four-fermion interaction, and therefore nonrenormalizable.
The ultraviolet divergences are handled by introducing a cutoff
which stays finite. Its functional form and numerical value,
therefore, are relevant for the predictions of the model.
Various cutoffs have been introduced, a good review of the
different possibilities and their numerical impacts
is given in \cite{Doring:1992sj}. Previously, the Schwinger proper-time 
cutoff was used almost exclusively until it was found
\cite{Diakonov:1996b} that this regularization violates
the momentum sum rule for the parton distributions, and
that Pauli-Villars regularization is favored in this respect.

The use of the Pauli-Villars regularization opens the possibility 
to introduce a numerical technique for computing the effective action
which has been developed previously 
\cite{Baacke:1990sb,Baacke:1992nh} and has been
applied to a various physical problems
involving fluctuation determinants \cite{Baacke:1994aj,Baacke:1995bk}.
It is based on using the Euclidean Green functions instead of
summation of levels. 
Computing functional determinants and
other expectation values involving the quark continuum by
summing over levels and using Minkowski space
wave functions requires the introduction of space
boundaries in order to discretize the spectrum. The associated
space cutoff has to be removed, introducing a numerical limiting
procedure. This limiting procedure seems to be technically
well under control, as can be seen, e.g., by comparing computations
of the sphaleron determinant using level summation 
\cite{Carson:1990rf} and using
Euclidean Green functions \cite{Baacke:1994aj}.
 Nevertheless, level summation is certainly not a 
very economical technique.  
At the same time an alternative computational
approach presents the possibility
to obtain the predictions of the model in an independent way.

The idea of replacing level summation by integrals over Euclidean Green
functions is actually a rather old one. It seems to go back
to Wichmann and Kroll \cite{Wichmann}. It has been used
extensively in calculations of the vacuum polarization in strong
fields \cite{Wichmann,Rafelski} and of the Casimir effect
\cite{Balian}.   A related way of encompassing the
discretization of the spectrum has been proposed by
Moussallam \cite{Moussallam:1989uk}, who uses Minkowski space 
phase shifts.

There are various methods to perform self-consistent computations
\cite{Reinhardt:1988}.
 As the equation of motion for the meson profiles 
requires to solve the equation of motion for the meson field,
or, equivalently, the chiral angle, it is advantageous to be able
to compute the functional derivative of the 
effective action with respect to the meson field. A technique for
computing such derivatives using Euclidean Green functions
has been set up recently \cite{Baacke:1995hw} and used for
a self-consistent computation of the bubble nucleation rate
in the electroweak theory \cite{Surig:1998ne}.
It is the purpose of this work to transfer these techniques,
{\sf mutatis mutandis}, to the NJL model. Apart from the
self-consistent computation of the mesonic profiles of the
nucleon we also formulate the sea quark contributions
to various other observables, such as the
moment of inertia, in terms of Euclidean Green functions.   
This latter aspect of our work is of course important;
without it, one would have to go back to level summation
in computing these observables and our technique would loose
much of its attractiveness.

The NJL model, as considered here, is not a unique theory; there are
many versions of it, the most elaborate ones 
\cite{Zuckert:1994}  include the
$\rho$, $a_1$ and $\omega$ vector mesons as well, as it was done 
before in the Skyrme model \cite{Lacombe:1986gi,Baacke:1987fg}.
 However, even within the
restricted class of models in which only the $\pi-\sigma$ fields
are taken into account, there is a wide variety. These 
models differ by the kind of regularization, as mentioned
above, but also by applying it to various parts of the
spectrum. In a renormalizable theory one would apply it only 
to the lowest perturbative contributions; in the NJL model
it is  applied to the finite parts as well. 
Furthermore, it was usually only applied to the quark sea , and not to
the valence contribution.
Some models also differ by the way in which the meson
degrees of freedom are varied: the variation can be extended to both
$\sigma$ and $\pi$ fields, or be restricted to the 
``chiral circle'', the 4-sphere being defined by $\sigma^2+\bpi^2=f_\pi^2$.   
Unfortunately, the existence of nucleon solutions is not a
robust property of the theory, such solutions are found
only within a small subset of the different versions, a situation
that must be considered as unsatisfactory.
We will not add to the ongoing discussion 
(see e.g. \cite{Golli:1998rf}), 
we will use that version of the model recently used by
Pobylitsa {\em et al.}\ \cite{Pobylitsa:1998tk} for computing 
the parton distributions. It only takes into account  the $\pi$ and $\sigma$ 
fields, which are varied on the chiral circle only, and
Paul-Villars regularization is applied
to the quark sea, not to the filled bound state.

\section{The model}
\setcounter{equation}{0}
Starting with the NJL-Lagrangian \cite{Nambu:1961a}
\be
{\lag}\sub{NJL}=
{\bar{\psi}}(i\gamma^{\mu}\partial_{\mu}-m)\psi+ \frac{G}{2}
\left[(\bar{\psi}\psi)^2+(\bar{\psi}i\gamma_5{\btau}\psi)^2\right]
\ee
one obtains after the standard bosonization procedure 
\be \label{boson}
S\sub{NJL}=\int  d ^4 x\Biggl\{ \bar{\psi}
\biggl[i\gamma^{\mu}\partial_{\mu}-g\Bigl(\sigma+i{\bpi}\cdot{\btau}
\gamma_5\Bigr)\biggr]
\psi-\frac{\mu^2}{2}\left(\sigma^2+{\bpi}^2\right)+
\frac{m\mu^2}{g}\sigma\Biggr\}\,.
\ee
The parameters of the model are the fermion self-coupling $G$, the quark 
mass $m$, and the cutoff scale $\Lambda$. In the bosonized 
version these parameters appear as the quark-meson coupling $g$ and the
symmetry breaking mass parameter $\mu$. They are related to the basic
parameters as
$g=\mu \sqrt{G}$ and $ m\mu^2=gf_\pi m_\pi^2$.
The latter equation expresses $\mu$ in terms of 
physical constants and of the coupling $g$ which remains a
free parameter. A further
relation is obtained from the gradient expansion of
the effective action. The resulting kinetic 
term of the pion field is normalized correctly if the 
Pauli-Villars cutoff is fixed as
\be
\Lambda=M\sqrt{\exp\left(\frac{4\pi^2}{N\sub{c} g^2}\right)}\,.
\label{lambda}
\ee
If the $\sigma-\pi$ field is varied only on the chiral circle, the second
term in Eq.\ (\ref{boson}) is absent.
For static pion fields the action is proportional to the time $\tau$;
 the two remaining parts of the effective action 
then contribute to the energy as
\bea
\label{SF}
E\sub{fer}&=&\frac{1}{\tau}\Tr\log\Bigl[{-i\gamma^{\mu}\partial_{\mu}
+\bM(\bx)}\Bigr]\,,\\
E\sub{br}&=&-{m_\pi^2f_\pi}\int d  ^3x \left(\sigma-f_\pi\right)\,.
\eea
Here the trace is taken over the quark sea and - in the case of baryons
 - over the filled bound states. $\bM(\bx)$ is given - on the chiral
circle -  by
\bea
\bM(\bx)&=&g\Bigl[\sigma+i\gamma_5{\btau}\cdot{\bpi}\Bigr]\nonumber\\
    &=&M\Bigl[\exp\left\{i\gamma_5{\btau}\cdot{\bphi}(\bx)
\right\}\Bigr]\nonumber\\
    &=&M\Bigl[\cos(|\bphi(\bx)|)
+i\gamma_5{\btau}\cdot{\hbphi}(\bx)\sin(|\bphi(\bx)|)\Bigr]\,,
\eea
where we have introduced the ``dynamical quark mass'' $M=g f_\pi$.
With the hedgehog ansatz ${\bphi}(\bx)=\hbx\vartheta(r)$ the mass becomes
\bea
\label{mass}
\bM(\bx)&=&M\Bigl[\cos(\vartheta(r))
+i\gamma_5{\btau}\cdot{\hbx}\sin(\vartheta(r))\Bigr]\pkt
\eea


\section{Basic relations}\label{basics}
\setcounter{equation}{0}
Given the profile $\vartheta(r)$, the energy of the corresponding nucleon state
consists of the symmetry breaking part that can be evaluated trivially, 
and the contributions of valence and sea quarks. In order to
evaluate the valence quark contribution we have to find the
bound state energy by solving, with appropriate boundary conditions,
 the Dirac equation
\be
(i\nu-H)\psi_0(\bx)=0
\pkt
\ee
Here we have introduced the Dirac Hamiltonian
\be\label{hami}
H=-i\balpha\cdot\bnabla+\gamma_0 \bM(\bx)
\pkt
\ee
The computation of the
quark sea contribution is more involved. We will recall here a
method introduced previously 
\cite{Baacke:1990sb} in which the
computation of the zero point energy is related to the Euclidean
Green function
\be \label{Greensum}
S\sub{E}(\bx,\bx',\nu)=\sum_{\alpha}\frac{\psi_\alpha(\bx)
\psi_\alpha^\dagger (\bx')}{-i\nu+E_\alpha}\pkt
\ee
The subscript $\alpha$ is a formal notation for the
discrete and continuum eigenstates
of the Dirac Hamiltonian $H$; we also indicate the positive energy
eigenstates by $\alpha>0$, the negative ones by
$\alpha<0$, and the valence eigenstate with $\alpha=0$.
$S\sub{E}$ satisfies the equation
\be\label{DGL1}
(i\nu -H)S\sub{E}(\bx,\bx',\nu)=-\delta^3(\bx-\bx')\,. 
\ee
The zero point energy
\be 
E_{\rm sea} = \sum_{\alpha < 0}E_\alpha
\ee 
can be computed as a contour integral around the positive imaginary
axis in the complex $\nu$-plane, see Fig. 1, 
as 
\be
E_{\rm sea} = \int_{C_-}  \frac{d\nu}{2\pi i}\nu 
\Tr \int d^3x S\sub{E}(\bx,\bx,\nu)
\pkt\ee

Deforming the contour to run along the real $\nu$ axis, and subtracting
the zero point energy of the free Dirac operator 
$H_0=-i\balpha\cdot\bnabla+\gamma_0M$, the integral
takes the form
\be
E_{\rm sea} = -i\int_{-\infty}^\infty \frac{d\nu}{2\pi}\nu 
\Tr \int d^3x \left[S\sub{E}(\bx,\bx,\nu)-S\sub{E,0}(\bx,\bx,\nu)
\right]\pkt\ee
The original contour, and therefore also the deformed one, takes
into account all negative energy states, whether they are
continuum or bound states. The only state that has to be 
considered separately is the positive energy bound state that
is occupied by the valence quarks and, therefore, should be included
as well. This can be done by deforming the contour to the one
presented in Fig. 1  as a dashed line. It is, however, 
more convenient to write the contibutions of this state separately, 
as we will do here.

It is convenient to introduce the bosonic Green function  $G\sub{E}$
via
 \be
   S\sub{E}=(i\nu+H)G\sub{E}\,
 \ee
which satisfies
\be\label{DGL2}
\left(\nu^2+H^2\right)G\sub{E}(\bx,\bx',\nu)=
\left[\nu^2-\Delta+M^2+{\cal{V}}(\bx)\right]
G\sub{E}(\bx,\bx',\nu)=\delta^3(\bx-\bx')
\ee
with the potential or vertex operator 
\be
  {\cal{V}(\bx)}=i\bgamma\cdot\bnabla\bM(\bx)\,.
\ee
In terms of $G\sub{E}$ the energy can be written as
\be
\label{zeropointbos}
E_0=\int_{0}^\infty \frac{d  \nu}{\pi}\nu^2
\int  d  ^3 x {\rm {Tr}}
\left[G\sub{E}(\bx,\bx,\nu)-G\sub{E,0}(\bx,\bx,\nu)\right]\,.
\ee
The expressions for the zero point energy and the subsequent manipulations 
are formal. Even after subtracting the free zero point energy, 
they only make sense if properly regularized.
In order to understand the divergences of the subtracted zero point 
energy we use the resolvent expansion of the Green function 
with respect to the potential 
${\cal V}$; if it is inserted into the expression for the zero
point energy we obtain
\be
\label{e0pert}
E_0=\sum_{n=1}^\infty{\frac{(-1)^n}{2n} \Tr \int \frac{d\nu}{2\pi}
\int \prod_{i=1}
^{n}{d^3x_iG\sub{E,0}(\bx_i-\bx_{i-1},\nu){\cal V}(\bx_i)}}
\ee
with $x_n=x_0$.
We have subtracted the free zero point energy by
omitting the zeroth order in the sum over $n$. The first order term
vanishes after taking the trace, and the only divergent term is the
second order one. Explicitly, it takes the form
\be
E_0^{(2)}=\frac{1}{4}\int\frac{d^3q}{(2\pi)^3}
\Tr \tilde{\cal V}(\bq)\tilde{\cal V}(-\bq)
\int_0^1 dx \int\frac{d^4p}{(2\pi)^4}
\frac{1}{\left[p^2+M^2+q^2x(1-x)\right]^2}
\kma
\ee
where $\tilde{\cal V}(\bq)$ is the Fourier transform 
of the potential.
The logarithmically divergent integral can be defined
by using, below the integrand, the Pauli-Villars subtraction
\be \label{E2reg}
E_0^{(2)}=\frac{1}{4}\int\frac{d^3q}{(2\pi)^3}
\Tr \tilde{\cal V}(\bq)\tilde{\cal V}(-\bq)
\int_0^1 dx \frac{1}{16 \pi^2}\left\{\ln \Lambda^2 -
\ln\left[M^2+q^2x(1-x)\right]\right\}
\pkt
\ee
Unlike in the case of renormalized perturbation theory here
the regularization will be extended over the finite contributions as well,
so that the regularized zero point energy reads
\be\label{regula}
E\sub{0,reg} = E\sub{0}(M)-\frac{M^2}{\Lambda^2}E\sub{0}(\Lambda)
\kma
\ee
where the subtraction is to be understood to be done below the
$\bx$ and $\nu$ integrals, respectively (see below), before
the partial wave summation and $\nu$ integration. 
The factor $M^2/\Lambda^2$ takes
into account that the potential $\tilde{\cal V}(\bq)$
contains a factor $M$, and a factor $\Lambda$ in the subtracted
part. These prefactors have to be compensated in order 
to ensure the cancellation of the divergent integrals. 

The perturbative expansion, Eq.\ (\ref{e0pert}),
can be used \cite{Baacke:1995hw} 
to obtain an expression for the derivative of
the zero point energy
\bea \nonumber
\frac{\delta E_0}{\delta\phi_a(\bz)}&=&
\sum_{n=1}^\infty {\frac{(-1)^n}{2} \int \frac{d\nu}{2\pi}}
\tr \int d^3x_1 
\frac{\delta {\cal V}(\bx_1)}{\delta \phi_a (\bz)} \\
&&\times\int\left( \prod_{i=2}
^{n}{d^3x_i G\sub{E,0}(\bx_i-\bx_{i-1},\nu){\cal V}(\bx_i)}
\right) G\sub{E,0}(\bx_1-\bx_n,\nu)\pkt\nonumber\\
\eea
The perturbative sum can be recollected into the
full nonperturbative Green function, so that
\bea
\label{deriSeff}
\frac{\delta E^{\overline{(2)}}_0}
{\delta{\phi}_a(\bz)}
=\tr\sum_{K,P}
\int d  ^3x\frac{\delta {\cal V}(\bx)}{\delta \phi_a (\bz)}
\int_{0}^\infty \frac{d  \nu}{2\pi}
G\sub{E}^{\overline{(1)}}(\bx,\bx,\nu)\pkt
\eea
Here and in the following we use superscripts $(j)$ to indicate
that the expression is of order $j$ in the potential
${\cal V}$, or its derivative. The symbol $\overline{(j)}$
indicates that the expression is evaluated to all orders in the 
potential starting with order $j$. So the exact zero point energy,
after subtracting the zeroth order, and taking account of the vanishing
of the first order, is of order $\overline{(2)}$, as indicated on the
l.h.s.\ of the last equation. $G\sub{E}^{\overline{(1)}}$ on the
r.h.s.\ includes all orders but the zeroth one.
It remains to evaluate the derivative of ${\cal V}({\bx})$ with
respect to $\phi_a$. One obtains
\bea
\frac{\delta {\cal V}
(\bx)}{\delta \phi_a(\bz)} &=& i M\bgamma\cdot\bnabla_x \delta^3(\bx-\bz)
\Biggl[i\gamma_5\tau_a
\exp (i\gamma_5\,\btau\cdot\bphi(\bx)) \\
&& +i\gamma_5 \left(\tau_a- \phi_a(\bx)\,\btau \cdot\hbphi (\bx)
 \right)\frac
{\sin|\bphi(\bx)|}{|\bphi(\bx)|}\Biggr]\,.\nonumber
\eea
Inserting the hedgehog ansatz, $\bphi(\bx)=\hbx \vartheta(r)$
and using the fact that within the trace the expectation value
of $\btau$ must be parallel to $\hbphi$, i.e.,
$\tau_a \to \hat x_a \btau\cdot \hbx$,
the derivative of the zero point energy takes the form 
\bea
\label{Seff31}
\left.\frac{\delta E_0^{\overline{(2)}}}{\delta{\phi}_a(\bz)}\right|
_{\hbphi(\bx)=\hbz\vartheta(r)}&=&
-M\tr\gamma_5 \tau_a \Bigl[\cos(\vartheta(r))-i\gamma_5\btau\cdot
{\hbz}\sin(\vartheta(r))\Bigr] \nonumber \\
&&\hspace{10mm}\times\int_{0}^\infty \frac{d  \nu}{2\pi}{\bgamma}
\cdot{\bnabla_z}G^{\overline{(1)}}\sub{E}(\bz,\bz,\nu)\,.
\eea
The gradient acts on the Green function at equal arguments. 
It is taken after carrying out that limit. 
The Euclidean Green function can be expanded \cite{Kahana:1984} 
with respect to K-spin harmonics $\Xi^{K,K_z}_n$ (for details see
Appendix A) as
\be
G\sub{E}(\bz,\bz',\nu)=\sum_{K,K_z,P}
g^{K,P}_{mn}(r,r',\nu) \Xi^{K,K_z}_m(\hbz)
\otimes\Xi^{K,K_z \dagger}_n(\hbz')
\pkt\ee
The radial Green functions form $ 4\times 4$ matrices; 
they can be written in terms of mode functions\footnote{
In the following we omit the K-spin 
and parity superscripts.}
 $f_n^{\alpha+}(\nu,r)$
and $f_n^{\alpha-}(\nu,r)$ which are solutions 
 regular at $r=0$ and as $r\to \infty$, respectively,
of a system of radial differential equations given in Appendix A.
Explicitly, they are given by
\be\label{gtheta}
g_{mn}(r,r',\nu)=\kappa\left[
            \theta(r-r')f^{\alpha +}_{m} (\nu,r)f^{\alpha -}_{n}(\nu,r') 
           +\theta(r'-r)f^{\alpha -}_{m} (\nu,r)f^{\alpha +}_{n}(\nu,r') 
            \right]\pkt
\ee
The superscript $\alpha$ labels $4$ linearly independent
solutions.

In this basis, and using the reduced Green functions, the zero point
energy takes the form {\cite{Baacke:1992nh}}
\be
\label{BaaE}
E_0^{\overline{(2)}}=N\sub{c}\int_0^\infty \frac{d\nu}{\pi}\nu^2\int 
 d  r\,r^2\sum_{K,P}(2K+1)
\left[g_{11}^{\overline{(2)}}+
g_{22}^{\overline{(2)}}+
g_{33}^{\overline{(2)}}+g_{44}^{\overline{(2)}}\right]\kma
\ee
where the Green functions are taken at $r=r'$. Analogously, the
functional derivative of the energy
is obtained as
\bea \label{dedphi}
&&\frac{\delta E_0^{\overline{(2)}}}{\delta{\phi}_a(\bz)}
\Biggr|_{\bphi(\bz)={\hbz}\vartheta(r)}=\nonumber
M\frac{N\sub{c}}{4\pi^2}{\hat z}_a\sum_{K,P}(-1)^K P
\\&&\hspace{1.5cm}
\times\int_0^\infty d  \nu \Biggl\{\sin(\vartheta(r))\Biggl[(2K+1)
\biggl(-g_{12}^{\overline{(1)}'}+g_{34}^{\overline{(1)}'}+\frac{2}{r}
\left\{-g_{12}^{\overline{(1)}}
+g_{34}^{\overline{(1)}}\right\}\biggr)\Biggr]\nonumber\\&&\hspace{1.5cm}
-\cos(\vartheta(r))\Biggl[\frac{1}{2}
\biggl(-g_{11}^{\overline{(1)}'}+g_{22}^{\overline{(1)}'}-
g_{33}^{\overline{(1)}'}
+g_{44}^{\overline{(1)}'}\biggr)\nonumber\\&&\hspace{2cm}+
\frac{1}{r}\biggl((K-1)g_{11}^{\overline{(1)}}+(K+1)g_{22}^{\overline{(1)}}+
K g_{33}^{\overline{(1)}}
+(K+2)g_{44}^{\overline{(1)}}\biggr)\nonumber\\&&\hspace{2cm}+
2\sqrt{K(K+1)}\biggl(-g_{14}^{\overline{(1)}'}-
g_{23}^{\overline{(1)}'}-\frac{1}{r}
\left\{3 g_{14}^{\overline{(1)}}
+g_{23}^{\overline{(1)}}\right\}\biggr)\Biggr]\Biggr\}\pkt
\eea
Both expressions have to be regulated as implied by Eq.\ (\ref{regula}).

The NJL-soliton is a system with baryon number equal to one. 
Therefore, one has to add the bound state part of
the fermionic energy
\be
  E_0\top{comp}=N\sub{c} E\top{bou}+E_0^{\overline{(2)}}\,.
\ee
The eigenvalue  equation for the bound state reads
\be
\label{EWgl}
\Bigl[-\Delta+{\cal V}^{0^+}(\bx)\Bigr]\psi_0=\omega_0^2\psi_0 \pkt
\ee
For $K^P=0^+$ the spinor $\psi_0$ is determined by
two radial wave functions $h_0(r)$ and $j_0(r)$, corresponding
to the components $u_3$ and $u_4$ for $K^P=0^+$ and for
 the bound state energy $E_0$; the potential
${\cal V}^{0+}$ is a $2\times 2$ matrix given in Appendix A.
The eigenfunctions are normalized as $\int|\psi_0(\bx)|^2 d^3x=1$.
Differentiating the bound state equation with respect to
$\phi_a(\bz)$ and projecting with $\psi_0^\dagger$ one finds
\bea
\frac{\delta}{\delta\phi_a (\bz)} \omega_0&=&
\frac{1}{2\omega_0}\int  d  ^3x
\psi_0^\dagger(\bx)\frac{\delta{\cal V}^{0^+}(\bx)}{\delta\phi_a(\bz)}
\psi_0(\bx)\,.
\eea
Inserting the explicit expressions 
for the potential and the eigenfunctions, the derivative of 
the bound state energy takes the form
\bea
\label{dedbou}
&&\frac{\delta E\top{bou}}{\delta \phi_a(\bz)}
\Biggr|_{\bphi(\bz)={\hbz}\vartheta(r)}
=\nonumber\\
&&\hspace{.8cm}M\frac{N\sub{c}}{4\pi\omega_0}{\hat z}_a\Biggl\{\sin(\vartheta(r))
\left[h_0'(r) j_0(r)+j_0'(r) h_0(r)+
\frac{2}{r}h_0(r)j_0(r)\right]
\nonumber\\&&\hspace{1.8cm}
-\cos(\vartheta(r))
\left[-h_0'(r) h_0(r)+j_0'(r)j_0(r)+\frac{2}{r}j_0^2(r)\right]
\Biggl\}\pkt
\eea
The bound state energy is convergent. Thus it does not 
need to be regularized. With a finite regulator, however, this
is subject to some arbitrariness. The same argument would hold
for any finite subset of sea quark states, or of all
the finite parts higher order terms
of the perturbative expansion.

Finally, the mesonic part of the energy and its derivative have
to be evaluated. This is straightforward, as they are given
by simple analytical expressions.
On the chiral circle one finds
\be
  E\top{br}=-4\pi m\sub{\pi}^2f\sub{\pi}^2\int d  r\,r^2
 \Bigl[\cos(\vartheta(r))-1\Bigr]\,
\ee
and
\bea
  \label{dedmes}
 \frac{\delta E\top{br}}{\delta{\phi}_a(\bz)}
\Biggr|_{\bphi(\bz)={\hbz}\vartheta(r)}&=&
 m\sub\pi^2f\sub{\pi}^2{\hat z}_a\sin(\vartheta(r))\,.
\eea


\section{Perturbative expansion of the Green function}
\setcounter{equation}{0}
After reduction of Eq.\ (\ref{DGL2}) to K-spin partial waves
(see Appendix A) the differential equation for the partial 
wave Green functions $g_{mn}(r,r',\nu)$ becomes
\bea
\label{GGleich}
&&\left[\delta_{nk}\left(\frac{d ^2}{d  r^2}+\frac{2}{r}
\frac{d }{d  r}-\frac{K_n(K_n+1)}{r^2}
-\kappa^2\right)-{\cal{V}}_{nk}(r)\right]
g_{km} ( r,r',\nu)\nonumber\\&&=
-\delta_{nm}\delta(r-r')/r^2 \,, 
\eea
where $\kappa=\sqrt{\nu^2+M^2}$ and where, again, we suppress the partial
wave indices $K$ and $P$. The potential ${\cal V}_{mn}$ depends
on the K-spin, its explicit form is given in Appendix A. 
As already mentioned, we use for the numerical
computation of the Green functions their standard expression
(\ref{gtheta}) in terms of mode functions $f_n^{\alpha\pm}(\nu,r)$.
The functions $f_n^{\alpha\pm}$ form $4\times 2$ linearly
independent systems (index $\alpha\pm$)
of $4$-component solutions (subscript $n$). 
A form independent of the choice of basis is given in
\cite{Baacke:1992nh,Baacke:1995hw}. Here we use a special convenient
basis; it is defined by splitting off the free solutions, i.e.\
the modified Bessel functions
$b^{+}_{K_n}(\kappa r)\equiv k_{K_n}(\kappa r)$ and
$b^{-}_{K_n}(\kappa r)\equiv i_{K_n}(\kappa r)$ 
via \footnote{ As for the mode functions
$f_n^{\alpha\pm}$ we omit the $K$ - spin and parity
assignment for the truncated mode functions 
$h_n^{\alpha\pm}$.}
\be
\label{fpm}
f^{\alpha \pm}_n(\nu,r)=\left[\delta^{\alpha \pm}_n+
h^{\alpha\pm}_n(\nu,r)\right]
                    b^{\pm}_{K_n}(\kappa r)
\ee
and by imposing the boundary condition
\be
  \lim_{r\rightarrow \infty}h^{\alpha\pm}_n(\nu,r)=0\pkt
\ee
The truncated mode-functions are obtained by solving the equations
\be
\label{DGLh}
 \left[
  \frac{d^2}{d  r^2}+2
    \left(\frac{1}{r}+\kappa\frac{b'^{\pm}_{K_n}(\kappa r)}
    {b^{\pm}_{K_n}(\kappa r)}
    \right) 
  \frac{d}{d r}
 \right]
 h_n^{\alpha\pm}(\nu,r)=
 {\cal{V}}^K_{nm}(r)
 \left[ \delta^\alpha_m+h_m^{\alpha\pm}(\nu,r)\right]
 \frac{b^{\pm}_{K_m}(\kappa r)}{b^{\pm}_{K_n}(\kappa r)}\,,
\ee
or, in a short form,
\be
  D h={\cal{V}}\left(1+h\right)\pkt
\ee
This equation can be used for a perturbative expansion.
Obviously, the functions $h_n^{\alpha\pm}(\nu,r)$ vanish to zeroth
order in ${\cal V}$, so, in the notation introduced in the
previous section, they are of order $\overline{(1)}$.
Once these solutions are known, the differential
equation may be iterated to obtain the contribution of
order $\overline{(2)}$ via
\bea 
  D h^{\overline{(1)}}&=&{\cal{V}}\left(1+h^{\overline{(1)}}\right)\,,\\
  \label{recur}
D h^{\overline{(2)}}&=&{\cal{V}}h^{\overline{(1)}}\,.
\eea
In terms of these functions the expression (\ref{gtheta}) 
becomes, for $r>r'$,
\bea
g_{nm}^{\overline{(1)}}(r,r',\nu)&=&\frac{\kappa}{2}
        \left[h_{n}^{\overline{(1)}m-}(\nu,r')+
       h_{m}^{\overline{(1)}n+}(\nu,r)+
       h_{n}^{\overline{(1)}\alpha-}(\nu,r')
       h_{m}^{\overline{(1)}\alpha+}(\nu,r)
        \right]
\nonumber \\ &&\times
k_{K_{m}}(\kappa r)i_{K_n}(\kappa r')\kma \\
g_{nm}^{\overline{(2)}}(r,r',\nu)&=&\frac{\kappa}{2}
       \left[h_{n}^{\overline{(2)}m-}(\nu,r')+
       h_{m}^{\overline{(2)}n+}(\nu,r)+
       h_{n}^{\overline{(1)}\alpha-}(\nu,r')
       h_{m}^{\overline{(1)}\alpha+}(\nu,r)
        \right]
\nonumber \\&& \times
k_{K_{m}}(\kappa r)i_{K_n}(\kappa r')
\pkt
\eea
These expressions are ready for being inserted into Eqs.~(\ref{BaaE}) and
(\ref{dedphi}). In renormalized perturbation theory one would, 
using further iterations,
reduce these expressions to order $\overline{(3)}$ and evaluate
the second order analytically as in Eq.~(\ref{E2reg}). 


\section{Quantization of collective coordinates}
\setcounter{equation}{0}
The hedgehog system is invariant under K-spin, i.e.,
under combined space and isospin rotations. For a 
rotating hedgehog state the Hamiltonian is modified
as \cite{Diakonov:1988}
 \bea
  H_\Omega(\bOmega)=  H(\bx)-\frac{1}{2}\bOmega\cdot\btau=
-i\balpha\cdot\bnabla+\gamma_0 \bM(\bx)-\frac{1}{2}\bOmega\cdot\btau\, ,
\eea
where $\bOmega$ is the angular velocity.
Assuming $\bOmega$ to be small, the problem can be treated perturbatively. 
The first order in $\bOmega$ vanishes, in second order one obtains
\be
  S\sub{eff}=-\tau\left[{E_0(\bOmega)+\frac 1 2 \Omega_a \theta_{ab}
 \Omega_b}\right]\,,
\ee
where $\theta_{ab}$ is the moment of inertia
\be
  \left.  \theta_{ab}=\frac{\delta^2 E_0(\bOmega)}
{\delta \Omega_a \delta \Omega_b}\right|_{\Omega=0}\,.
\ee
$\theta_{ab}$ is proportional to the unit matrix, $\theta_{ab}=
\theta \delta_{ab}$.
The collective coordinate can be quantized in the usual way,
leading to an extra term $J(J+1)/2\theta$ in the energy.

Taking the second derivative with respect to $\Omega$
of the fermion Green function
\be
S\sub{E}^\Omega(\bx,\bx',\nu)=-\langle x|\frac{1}{i\nu -H_\Omega(\bOmega)} 
|x' \rangle
\ee
one obtains
\bea
  \frac{\delta^2 S^\Omega\sub{E}(\bx,\bx',\nu)}{\delta \Omega_a\delta \Omega_b}
&=&-\langle x|\frac{1}{i\nu -H_\Omega(\bOmega)}\frac{\tau_a}{2}\frac{1}
{i\nu -H_\Omega(\bOmega)} 
\frac{\tau_b}{2}\frac{1}{i\nu -H_\Omega(\bOmega)}|x' \rangle \nonumber\\
&&-\langle x|\frac{1}{i\nu -H_\Omega(\bOmega)}\frac{\tau_b}{2}\frac{1}
{i\nu -H_\Omega(\bOmega)}
 \frac{\tau_a}{2}\frac{1}{i\nu -H_\Omega(\bOmega)}|x' \rangle \nonumber\\
\\
&=&-\frac{i}{4}\frac{\partial}{\partial \nu}\langle x| \tau_a\frac{1}
{i\nu -H_\Omega(\bOmega)} \tau_b\frac{1}{i\nu -H_\Omega(\bOmega)}|x' \rangle
\,.
\eea
Inserting this equation in Eq.\ (\ref{zeropointbos})
the tensor can be calculated via
\be
\label{tm}
  \left.\frac{\delta^2 E_0}{\delta \Omega_a\delta \Omega_b}\right|_{\bOmegaA=0}
=N\sub{c}\int_{-\infty}^\infty \frac{d \nu}{8\pi}\int  d  ^3 x 
\int d^3 x' {\rm {tr}}
 \left[\tau_a S\sub{E}^0(\bx,\bx',\nu)\tau_b S\sub{E}^0(\bx',\bx,\nu)
  \right]\, ,\nonumber\\
\ee
where $S\sub{E}^0(\bx,\bx',\nu)$ is defined by Eq.\ (\ref{DGL1}).
This expression can be rewritten with the "bosonic" Green function (\ref{DGL2})
\bea
\theta_{ab}&=&N\sub{c}\int_{-\infty}^\infty \frac{d\nu}{8\pi}
\int  d  ^3 x \int d^3 x'\\ && \hspace{1cm}\times{\rm {tr}}
 \Bigl[\tau_a \left\{i\nu+H(\bx)\right\}G\sub{E}(\bx,\bx',\nu)\tau_b
\left\{i\nu+H(\bx')\right\}G\sub{E}(\bx',\bx,\nu)  \Bigr]\,.\nonumber
\eea
The bound state is occupied and, therefore, included into the negative
continuum by choosing the dashed integration contour displayed
in Fig. 1. If the Green function is expanded into eigenfunctions 
of the Hamiltonian as in Eq.~(\ref{Greensum}), it can be readily verified 
that this way one obtains transitions between the positive and 
negative continuum states as well as transitions between the
bound state and the positive continuum as discussed in \cite{Goeke:1991fk}.
Here this detailed structure is not explicit. 

As for the energy, we will consider the
bound state contribution separately. We decompose the dashed contour
into a contour running along the real $\nu$ axis and a small circle
around the bound state pole. We will denote the former contribution
which describes the continuum-continuum-transitions by a superscript
$c-c$, and the bound state contributions by the superscript $b-c$ as
it involves matrix elements between the bound state and the continuum.  
We begin with considering the continuum contributions.
 
The partial wave reduction of the Green function
$G\sub E (\bx,\bx',\nu)$ has been introduced above and is discussed in
Appendix A. Here we need the partial wave reduction for the
fermion propagator $S\sub E=(i\nu +H)G\sub E$; we denote the associated
matrix elements by $s^{K}_{ m n}(r,r',\nu)$.
 This Green function can again be decomposed into mode functions
as in Eq.~\ref{gtheta}, one just has to replace the functions
$f_n(\nu,r)$ by the fermionic mode functions $u_{n}(\nu,r)$. These 
are given explicitly in Appendix A. Furthermore, we have to take
the trace with isospin matrices $\tau_a$. As $\theta_{ab}
\propto \delta_{ab}$ it is sufficient to compute
$\theta_{33}$. The action of $\tau_3$ on the
K-spin harmonics is given in Appendix B.
Using these expressions, the angular integration 
over $d\Omega_{r}$ and $d\Omega_{r'}$ 
and the summation over the third component of $K$ and $K'$ can be
performed. $K'$ is fixed by the angular 
integration to the values $K$, $K-1$ or $ K+1$.
Finally, one obtains
\begin{eqnarray}\nonumber
&&\int d\Omega_{r}\int d\Omega_{r'}
{\rm {tr}} \Bigl[\tau_3 \left\{i\nu+H(\bx)\right\}G\sub{E}(\bx,\bx',\nu)
\tau_3\left\{i\nu+H(\bx')\right\}G\sub{E}(\bx',\bx,\nu)  \Bigr]=
\\\nonumber&&
\hspace{0cm}\sum_{P,K=1}^\infty \Biggl[\frac{(K+1)(2K+1)}{3K}\biggl\{
  { s^{K}_{ 11}}({r}, { r'},\nu ){ s^{K}_{ 11}}({ r'}, {r},\nu)
+ { s^{K}_{ 22}}({r}, { r'},\nu){ s^{K}_{22}}({ r'}, {r},\nu)
\\\nonumber&&
\hspace{4.8cm}+{s^{K}_{ 12}}({r}, { r'},\nu){ s^{K}_{ 21}}({ r'}, {r},\nu)
+ { s^{K}_{ 21}}({r}, { r'},\nu){ s^{K}_{ 12}}({ r'}, {r},\nu)
\biggr\}\\\nonumber&&
\hspace{1.8cm}+\frac{(2K+1)K}{3(K+1)}\biggl\{
  { s^{K}_{ 33}}({r}, { r'},\nu){ s^{K}_{ 33}}({ r'}, {r},\nu)
+ { s^{K}_{ 44}}({r}, { r'},\nu){ s^{K}_{ 44}}({ r'}, {r},\nu)
\\\nonumber&&
\hspace{4.8cm}+ {s^{K}_{34}}({r}, { r'},\nu){ s^{K}_{ 43}}({ r'},{r},\nu)
+ { s^{K}_{ 43}}({r}, { r'},\nu){ s^{K}_{ 34}}({ r'}, {r},\nu)
\biggr\}
\\&&\hspace{1.8cm}\nonumber
-\frac{(2K+1)}{3}\biggl\{
  { s^{K}_{ 14}}({r}, { r'},\nu){ s^{K}_{ 41}}({ r'}, {r},\nu)
+ { s^{K}_{ 41}}({r}, { r'},\nu){ s^{K}_{ 14}}({ r'}, {r},\nu)
\\\nonumber&&
\hspace{4.8cm}+ { s^{K}_{ 13}}({r}, { r'},\nu){ s^{K}_{31}}({r'},{r},\nu)
+ { s^{K}_{31}}({r}, { r'},\nu){ s^{K}_{13}}({ r'}, {r},\nu)
\\\nonumber&&
\hspace{4.8cm}+ { s^{K}_{32}}({r}, { r'},\nu){ s^{K}_{ 23}}({r'},{r},\nu)
+ { s^{K}_{23}}({r}, { r'},\nu){ s^{K}_{ 32}}({ r'}, {r},\nu)
\\\nonumber&&
\hspace{4.8cm}+ { s^{K}_{ 24}}({r}, { r'},\nu){ s^{K}_{ 42}}({r'},{r},\nu)
+ { s^{K}_{ 42}}({r}, { r'},\nu){ s^{K}_{24}}({ r'}, {r},\nu)
\biggr\}
\\\nonumber&&\hspace{1.8cm}
+\frac{4K^2-1}{3K}\biggl\{
  { s^{K}_{ 22}}({r}, { r'},\nu){ s^{K-1}_{ 44}}({ r'}, {r},\nu)
+ { s^{K}_{ 12}}({r},{ r'},\nu){ s^{K-1}_{ 43}}({ r'}, {r},\nu)
\\\nonumber&&
\hspace{3.8cm}+ {s^{K}_{21}}({r}, {r'},\nu){ s^{K-1}_{34}}({r'},{r},\nu)
+ { s^{K}_{ 11}}({r}, { r'},\nu){ s^{K-1}_{33}}({ r'}, {r},\nu)
\\\nonumber&&
\hspace{3.8cm}+{s^{K-1}_{33}}({r},{ r'},\nu){ s^{K}_{11}}({r'},{r},\nu)
+ { s^{K-1}_{ 44}}({r},{r'},\nu){ s^{K}_{22}}({ r'}, {r},\nu)
\\\nonumber&&
\hspace{3.8cm}+ {s^{K-1}_{34}}({r},{ r'},\nu){ s^{K}_{ 21}}({r'},{r},\nu)
+ { s^{K-1}_{ 43}}({r},{r'},\nu){ s^{K}_{12}}({ r'}, {r},\nu)\biggr\}
\Biggr]\\ \,.
 \end{eqnarray}
As we need this expression in order $\overline{(2)}$, the
modified Green functions have to be inserted in order
$\overline{(1)}$. One can take advantage of the factorization
of the Green functions into mode functions,
see~\ref{gthetatilde}, to rewrite this expression in the form
 \bea\label{thetaH}
&& \theta\top{c-c}=\left.\frac{\delta^2 E_0}
{\delta \Omega_a\delta \Omega_b}\right|_{\bOmegaA=0}
=N\sub{c}\int_{-\infty}^\infty \frac{d  \nu}{4\pi} \sum_{P,K=1}^\infty
\int_0^\infty  d  r r^2\int_0^r  d  r' r'^2 
\nonumber\\&&\hspace{1cm}\Biggl[
H^{+\alpha\beta}_{1K}(\nu,r)\left\{\frac{(K+1)(2K+1)}{3K}
H^{-\beta\alpha}_{1K}(\nu,r')-
\frac{2K+1}{3}H^{-\beta\alpha}_{2K}(\nu,r')\right\}\nonumber\\&&\hspace{1.2cm}+
H^{+\alpha\beta}_{2K}(\nu,r)\left\{\frac{K(2K+1)}{3(K+1)}
H^{-\beta\alpha}_{2K}(\nu,r')-
\frac{2K+1}{3}H^{-\beta\alpha}_{1K}(\nu,r')\right\}\nonumber\\&&\hspace{1.2cm}+
H^{+\alpha\beta}_{3K}(\nu,r)\left\{\frac{4K^2-1}{3K}
H^{-\beta\alpha}_{4K}(\nu,r')\right\}\nonumber\\&&\hspace{1.2cm}
+H^{+\alpha\beta}_{4K}(\nu,r)\left\{\frac{4K^2-1}{3K}
H^{-\beta\alpha}_{3K}(\nu,r')\right\}\Biggr]\,.
 \eea
The functions $ H^{\pm\alpha\beta}_{iK}(\nu,r)$ are defined as
 \bea
   H^{\pm\alpha\beta}_{1K}(\nu,r)&=&\kappa\left[u_{ 1,K}^{\alpha\pm }(\nu,r)
f_{1,K}^{\beta\pm}(\nu,r)+ u_{ 2,K}^{\alpha\pm}(\nu,r)
f_{2,K}^{\beta\pm}(\nu,r)\right]\,,\\
   H^{\pm\alpha\beta}_{2K}(\nu,r)&=&\kappa\left[u_{ 3,K}^{\alpha\pm}(\nu,r)
f_{3,K}^{\beta\pm}(\nu,r)+u_{ 4,K}^{\alpha\pm}(\nu,r)
f_{4,K}^{\beta\pm}(\nu,r)\right]\,,\\
   H^{\pm\alpha\beta}_{3K}(\nu,r)&=&\kappa\left[u_{ 1,K}^{\alpha\pm}(\nu,r)
f_{3,K-1}^{\beta\pm}(\nu,r)+u_{ 2,K}^{\alpha\pm}(\nu,r)
f_{4,K-1}^{\beta\pm}(\nu,r)\right]\,,\\
   H^{\pm\alpha\beta}_{4K}(\nu,r)&=&\kappa\left[u_{ 3,K-1}^{\alpha\pm}(\nu,r)
f_{1,K}^{\beta\pm}(\nu,r)+u_{4,K-1}^{\alpha\pm}(\nu,r)
f_{2,K}^{\beta\pm}(\nu,r)\right]\,,
\eea
in terms of the mode functions $f_n^{\alpha\pm}$ and
$u_{\tilde n}^{\alpha\pm}$ defined in Appendix A.
One has to combine the orders of 
$H^{\pm\alpha\beta}_{i}$ in such a way that the result is 
of total order $\overline{(2)}$:
 \bea
\theta_{ab}\top{c-c\overline{(1)}}&\sim&H^{+{\overline{(1)}}}
H^{-{\overline{(1)}}}+H^{+{(0)}}H^{-{\overline{(1)}}}+
H^{+{\overline{(1)}}}H^{-{(0)}}\,.
 \eea
In fact the first order part which has been 
included on the right hand side for practical convenience
vanishes. The functions $H^{\pm\overline{(1)}}$ are 
obtained as
\bea
H^{\pm{\overline{(1)}}}&\sim&f^{\pm{\overline{(1)}}}f^{\pm{\overline{(1)}}}+
f^{\pm{(0)}}f^{\pm{\overline{(1)}}}+
f^{\pm{\overline{(1)}}}f^{\pm{(0)}}\,;
 \eea
the functions $H^{\pm (0)}$ are composed of free Bessel functions.

Since the imaginary part of the integral of Eq.\ (\ref{thetaH}) 
is antisymmetric, the result equals twice the real part, 
integrated from $\nu = 0$ up to $\infty$.
It is implied that the expressions have to be regularized by
 Pauli-Villars subtractions. 

Having presented the continuum-continuum contributions to
the moment of inertia we now turn to the bound state contribution.
This contribution is given, in terms of eigenfunctions
of the Dirac operator by 
\cite{Goeke:1991fk,Wakamatsu:1991}
\bea
\left.\frac{\partial^2\omega_0}{\partial \Omega_a\partial \Omega_a}
\right|_{\bOmegaA=0}
&=&\frac{N\sub{c}}{2}\sum_{m\ne {\rm bou}}\frac{\langle\psi_0|\tau_a|
\psi_m\rangle\langle\psi_m|\tau_b|\psi_0\rangle}
{E^m-E\top{bou}}\,.
\eea
Using the (\ref{Greensum}) for the Green function we find 
that this expression for the moment of inertia is identical to
\bea
\theta\top{b-c}_{ab}=\left.\frac{\partial^2\omega_0}
{\partial \Omega_a\partial \Omega_a}\right|_{\bOmegaA=0}
&=&\frac{N\sub{c}}{2}{\rm {tr}}
\int d^3x d^3x'\psi_0(\bx)\psi^\dagger_0(\bx')
\tau_a S(\bx',\bx,-iE\top{bou})\tau_b\,.\nonumber\\
\eea
The Euclidean Green function at any imaginary argument can again be related to
the bosonic Green function via 
\be
  S\sub{E}(\bx,\bx',-iE\top{bou})=
(H+E\top{bou})G\sub{E}(\bx,\bx',iE\top{bou})\,.
\ee
Since the energy of the bound state is smaller than the mass $M$, 
the calculation of the Green function is analogous to the
one for the continuum part, Eq. (\ref{thetaH}), with
\be
  \kappa^2=M^2-E\sub{bou}^2\,.
\ee
Furthermore, the valence state $0^+$ only couples to $K^P=1^+$ continuum
states. Therefore, the expression for this contribution 
reduces to
\bea
\theta\top{b-c}_{ab}&=&\frac{N\sub{c}}
{2\kappa} \int_0^\infty  d  r r^2\Biggl[
{\cal{H}}^{+\alpha}_{3}(\nu,r)\int_0^r d r' r'^2{\cal{H}}^{-\alpha}_{4}(\nu,r')
\nonumber\\&&\hspace{1cm}+
{\cal{H}}^{+\alpha}_{4}(\nu,r)\int_0^r d r' r'^2{\cal{H}}^{-\alpha}_{3}(\nu,r')\Biggr]
\eea
with
 \bea
 {\cal{H}}^{\pm\alpha}_{3}(\nu,r)&=&\frac{\kappa}{2E\sub{bou}}
\left[u_{1,1}^{\alpha\pm}(\nu,r)h_0(r)
+u_{2,1}^{\alpha\pm}(\nu,r)j_0(r)\right]\,,\\
  {\cal{H}}^{\pm\alpha}_{4}(\nu,r)
&=&\kappa\left[f_{1,1}^{\alpha\pm}(\nu,r)h_0(r)+ 
f_{2,1}^{\alpha\pm}(\nu,r)j_0(r)\right]\,.
\eea
In the first of these equations we have used the
Dirac equation to relate the bosonized wave functions
$f_{i,0}$ of the bound state to the fermionic components 
$h_0$ and $j_0$.

As for the bound state contribution to the energy, this part of the
moment of inertia is finite and will not be Pauli-Villars subtracted.
Adding the $c-c$ and $b-c$ contributions
 the moment of inertia is given by
\be
  \theta=\theta\top{b-c}+\theta\top{c-c}(M)-
\frac{M^2}{\Lambda^2}\theta\top{c-c}(\Lambda)\,.
\ee
Proceeding in an analogous way one can obtain expressions for
the expectation values of other observables as well.
Expressions for
$\langle\Sigma_3\rangle$ and $\langle L_3\rangle$ in terms
of mode sums have been derived in \cite{Wakamatsu:1991}.
The spin expectation value is given by
\be
\langle\Sigma_3\rangle=
-\frac{1}{\theta}\frac{N\sub{c}}{2} \sum_{m,n} 
\frac{\langle \psi_n|\tau_3|\psi_m 
\rangle \langle \psi_m|\Sigma_3|\psi_n \rangle}{E_m-E_n}
\pkt
\ee
It can be rewritten in terms of the Euclidean Green function as
\be
\langle \Sigma_3\rangle
=-\frac{N\sub{c}}{\theta}\int_{-\infty}^\infty \frac{d \nu}{8\pi}\int  d  ^3 x 
\int d^3 x' {\rm {tr}}
 \left[\tau_3 S\sub{E}^0(\bx,\bx',\nu)\sigma_3 S\sub{E}^0(\bx',\bx,\nu)
  \right]
\ee
where the contour again includes the bound state. The bound state
and continuum contributions are obtained separately, as above.


\section{Observables}
\setcounter{equation}{0}
The expressions for the energy and for the moment of inertia have 
already been presented. 
The mass of the low-lying baryons with K-spin 0 is given by
the sum of static and rotational energy as
\be
M_{J}=E_0+\frac{J(J+1)}{2\theta}\,,
\ee
therefore
\be
  \label{massnuk}
  M\sub{N}=E_0+\frac{3}{8\theta}\ee
and
\be
  \label{massdiff}
  M_\Delta-M\sub{N}=\frac{15}{8\theta}-\frac{3}{8\theta}=\frac{3}{2\theta}\,.
\ee
The nucleon sigma term is defined as
\be
  \Sigma=m_0\int  d  ^3 x \langle\overline{q}q\rangle\,.
\ee
As shown in \cite{Meissner:1990} it is given simply by the 
symmetry breaking part of the energy as
\be
  \Sigma=E\top{br}\,.
\ee
The experimental value is $45\pm9$ MeV \cite{Gasser:1991ce}.
The pion-nucleon coupling constant can be obtained
\cite{Adkins:1983} from the long range behavior of the
meson profile 
\be
  C=f_\pi \lim_{r\rightarrow \infty} r^2  \sin(\vartheta)\frac{\exp(m_\pi r)}
{1+m_\pi r}
\ee
as
\be
  \label{gacont}
  g\sub{\pi NN}=\frac{8}{3}\pi M\sub{N} C\,.
\ee
The axial vector coupling constant is given 
\cite{Wakamatsu:1990ag,Meissner:1991} by
the expectation
value
\be
g\sub{A}=\langle p\uparrow|\gamma_0\tau_3\gamma_3\gamma_5|p\uparrow\rangle
\pkt\ee
It consists of a valence state contribution
\bea \label{ga_val}
  g\sub{A}\top{bou}&=&-\frac{N\sub{c}}{3}\tr\int  d  ^3 x 
\left(\gamma_0\gamma_3\gamma_5\tau_3\right)\psi_0(\bx)
\psi^\dagger_0(\bx)
\nonumber
\\&=&\frac{N\sub{c}}{3}\int  d  r r^2\left[h^2(r)-\frac{1}{3}j^2(r)\right]
\eea
and a continuum part which, using the Euclidean Green function,
can be written as
\bea \label{ga_sea}
  g\sub{A}\top{con}&=&-\frac{N\sub{c}}{3}\tr\int  d  ^3 x\int_{-\infty}^\infty 
\frac{d  \nu}{2\pi} 
\left(\gamma_0\gamma_3\gamma_5\tau_3\right)S\sub{E}(\bx,\bx,\nu)
\nonumber
\\&=&
-\frac{2}{9}N\sub{c}\int_0^\infty \frac{d  \nu}{2\pi}\int_0^\infty  d  r r^2
\sum_{K^P}
\left[
  \left(2K+1\right)s_{11}(r,r,\nu)
 - \left(2K-1\right)s_{22}(r,r,\nu)\nonumber\right.
\\&&\left.\hspace{4.2cm}
 - \left(2K+3\right)s_{33}(r,r,\nu)
 + \left(2K+1\right)s_{44}(r,r,\nu)\nonumber\right.
\\&&\left.\hspace{4.2cm}
 +4\sqrt{K\left(K+1\right)}
 \left\{s_{23}(r,r,\nu)+s_{32}(r,r,\nu)   
 \right\}
\right]\,.\nonumber\\
\eea
Pauli-Villars subtraction is implied.
With the Goldberger-Treiman relation the axial vector coupling constant
can also be calculated via
\be
g\sub{A}\top{G-T}=\frac{f_\pi}{M\sub{N}}g\sub{\pi NN}
\pkt
\ee
The quadratic radius of
\bea
  \label{rhoch2}
  \langle R^2 \rangle\sub{bou}&=&\int_0^\infty r^4  d  r\left\{
h_0^2(r)+j_0^2(r)
\right\}\,,\\  \langle R^2 \rangle\sub{con}&=&\frac{-1}{2\pi}\int_0^\infty  d
 \nu \int_0^\infty r^4 d r\sum_{K^P}\left\{s_{11}+
s_{22}+s_{33}+s_{44}\right\}
\eea
has to be compared with an experimental value of $0.62$ fm$^2$. The 
continuum part is convergent, but so small that regularizing the integral 
does not change the result.


\section{Numerics}
\setcounter{equation}{0}
We have numerically implemented the expressions for the energy and its
functional derivative presented in section 3 in the way described
in \cite{Baacke:1992nh} for the energy, and in \cite{Baacke:1995hw} for its
functional  derivative. 

The iteration proceeds as follows: For a given meson profile
$\vartheta(r)$ one computes the mode functions
and evaluates the functional derivative of the energy.
One then requires the vanishing of the functional derivative,
\be
  \label{eom}
  {\hat z}_a
\left(
\frac{\delta E^{\overline{(2)}}_{0,M}}{\delta{\phi}_a(\bz)}
+\frac{\delta E\top{bou}}{\delta{\phi}_a(\bz)}
+\frac{\delta E\top{br}}{\delta{\phi}_a(\bz)}
-\frac{M^2}{\Lambda^2}\frac{\delta E^{\overline{(2)}}_{0,\Lambda}}
{\delta{\phi}_a(\bz)}
\right)
_{\bphi(\bz)={\hbz}\vartheta(r)}=0 \pkt
\ee
As can be seen from Eqs. (\ref{dedphi}), (\ref{dedbou})
 and (\ref{dedmes}), this equation takes
the form
\be
\frac{\delta E}{\delta{\phi}_a(\bz)}\Biggr|
_{\bphi(\bz)={\hbz}\vartheta(r)} =\hat {z}_a \left[A(r)
\cos(\vartheta(r))+B(r)\sin(\vartheta(r))\right]\pkt
\ee 
The coefficient functions $A(r)$ and $B(r)$ are the results of the
numerical computation.
Extremizing the energy by requiring the functional 
derivative to vanish fixes $\vartheta(r)$ via
$\tan(\vartheta(r))=-A(r)/B(r)$. This profile
is used as the input for the next iteration.
This method of iteration has been used previously by 
\cite{Meissner:1991}.

The functions $h^{\alpha\pm}(\nu,r)$ have been computed in
order $\overline{(1)}$ and $\overline{(2)}$
by solving (\ref{DGLh}) and its recursion (\ref{recur})
using a Runge-Kutta scheme.
The accuracy of these solutions was checked by using the
Wronskian relation, which was constant to at least $6$ significant
units. The sum over the K-spin was extended to 
$K_{\rm max}=16$ during the iteration
and to $K_{\rm max}=20$ for the final result. It is straightforward
to derive, e.g., using asymptotic expansions
for the lowest order perturbative contributions,
 that the power behaviour
in angular momentum should be, after regularization,
as $K^{-3}$.  In Fig. 2a we show the terms in the sum over 
angular momenta for the integrand of the energy at $\nu=1$
with a power fit $A K^{-3}+B K^{-4}$. An analogous
example is given for the $\nu$ integrand of $g_A$,
Eq. (\ref{ga_sea}) at $\nu=0$.
Using this power fit it is then straightforward to include 
the sum above $K_{\rm max}$. So with a very good approximation
the angular momentum sum runs, effectively, up to $K=\infty$.
Likewise the integral over $\nu$ can be extended to
$\nu = \infty$ by using a power fit. The $\nu$ integrand
for the energy is displayed in Fig. 2c. Here the expected 
power behaviour is $\nu^{-2}$ and, based on the power fit
displayed in this figure we have appended the integral
above  $\nu = 5 M$.

The numerical results for the energy and other static
parameters are presented in Table \ref{tab1} and in 
Figs. 3 - 7.


\section{Results and conclusions}
\setcounter{equation}{0}
We have presented here a self-consistent computation of the 
nucleon ground state in the Nambu-Jona-Lasinio model.
In contrast to most previous calculations we have used a 
Pauli-Villars cutoff. It has been shown recently
\cite{Diakonov:1996b} that such a cutoff 
is favored by parton sum rules. As the main object of this work
we have introduced an alternative method of numerical computation,
based on Euclidean Green functions instead of
the use of the quark eigenfunctions for real energies.
Our method has the advantage that it is not necessary 
to discretize the continuous spectrum of the sea quarks.
While this discretization and the associated limiting
procedure seem to be well under
control, finite boundaries may introduce spurious effects,
as discussed in \cite{Wakamatsu:1991}.
Although our results essentially confirm those of other groups, 
this agreement is by no means guaranteed. 

Besides presenting the analytical framework for the computation 
of self-consistent profiles, based on explicit expressions for
the energy and its functional derivative, we have also derived 
explicit expressions for other observables. Again the quark sea
contributions can be formulated in terms of the Euclidean Green
function.  

Our numerical results are presented in Table \ref{tab1}
 and plotted in Figs. 3 to 7.
In Table \ref{tab1} we also give some results obtained for $g=4$ in Ref. 
\cite{Doring:1992sj}, when using the same regularization.
In view of the difference of the numerical approaches the agreement 
is very satisfactory. This agreement holds as well for the mesonic profiles
$\pi(r)$, plotted in  Fig. 3. In Figs. 4
- 7 we plot the various parts 
of the energy, the axial vector coupling
$g_A$, the mesonic profiles $\vartheta(r)$ and the moment of inertia,
as functions of the coupling $g$. One sees that a value of $g \simeq 4$
is preferred by the comparison with experiment.

The nucleon mass is still too high, but 
somewhat lower than the one
obtained with Schwinger proper-time cutoff. A lower value
for $M\sub{N}$ can be obtained by minimizing the sum of 
fluctuation energy and rotational energy \cite{Schleif:1997}.
We have not followed this issue here. It can be inferred
from multiparticle dynamics (see, e.g. \cite{Ring:1980}) 
that one should subtract
an energy corresponding to the c.m.~motion of the quarks,
 which would
result in a the mass near to the physical value.
For the present selfconsistent condensate-quark
state there is, however, no compelling proof for
such a procedure. So such a subtraction is 
performed by some authors \cite{Pobylitsa:1992bk}, 
but has not become a general standard. 
  
A major difference to other publications
on the chiral quark model is observed for the axial coupling
constant $g\sub{A}$ which is computed from the quark current, 
and for which we obtain values between $1.15$ 
and $1.37$ as shown in Fig. 5. 
Most other authors have used the
Schwinger proper-time cutoff, they obtain
results well below $1$.
However, similar values for $g\sub{A}$ have been obtained recently
 by Golli {\em et al.} \cite{Golli:1998rf} who use a
Gaussian cutoff for the effective quark mass.
The bound state contribution agrees with the one given, e.g., 
in \cite{Doring:1992sj}. The major part of the increase
of $g\sub{A}$ with $g$ comes from the continuum part.
The values for $g\sub{A}$ obtained from the asymptotic
behavior of the soliton profile via the Goldberger-Treiman
relation are somewhat lower than those computed directly
from the quark currents, they are in the range between
$1.15$ and $1.17$. While a small violation of the Goldberger-Treiman
relation is expected, the appreciable increase of this violation
with $g$ is somewhat troublesome.

As to the variation of the axial vector coupling with
the coupling $g$, we find a monotonous increase.
The bound state contribution decreases with $g$, this
is, however, overcompensated by an increasing
continum contribution.
This trend is different from the one found by other
authors (see, e.g., \cite{Doring:1992sj,Wakamatsu:1993up}),
 mostly using the Schwinger proper-time cutoff.
 It agrees, however,
with the one found in \cite{Doring:1992sj} for their
 Pauli-Villars I cutoff, as we may infer from
the two values  $g\sub{A}=.94$ at  $g=3.85$ and 
$g\sub{A}=.96$ at $g=4$. The dependence of $g\sub{A}$ on $g$ 
has again the same trend as ours in Ref. \cite{Golli:1998rf}
with the Gaussian mass cutoff\footnote{The correspondence
between their parameters and ours is somewhat involved,
in the range considered here
 their values of $1/G$ are monotonuos with
our coupling $g$, as are their and our values for
the nucleon mass.} 
In the same way the trend of the 
isoscalar quadratic radius $\langle R^2 \rangle$ 
with $g$ is found opposite to the one in other 
cutoff schemes; it again agrees with the
Pauli-Villars I results of \cite{Doring:1992sj}.
So it seems that it is the regularization scheme
and not some numerical deficiency which
causes the differences in the trends, an unsatisfactory 
situation which requires further investigation.

While the absolute value of $g\sub{A}$ obtained here suggests a satisfactory
agreement with experiment (see, e.g., \cite{Matsinos:1998wp})
 it has to be taken into consideration
that $g\sub{A}$ is only calculated in ${\cal O}(\Omega^0)$. As shown in 
\cite{Wakamatsu:1993up} the next orders in $\Omega$ lead to 
additional contributions of  $g^1\sub{A}\sim 0.4$ and $g^2\sub{A}\sim 0.2$,
if computed with Schwinger proper-time regularisation.
Thus in fact $g\sub{A}$ is overestimated here, as 
in nearly all regularization schemes,
as discussed in \cite{Wakamatsu:1993up}.
The problem is not resolved entirely, however, as
apparently the expansion in $\Omega$ does not
converge well, so that higher corrections may still modify the
results appreciably.

In conclusion we have presented here a new approach to
computing self-consistent meson profiles and
static  observables of the nucleon in the 
Nambu-Jona-Lasinio model, using a Pauli-Villars cutoff.
The agreement with previous
analyses using the same cutoff but different numerical
methods is satisfactory in general, some
results given here are new. In view of the fact that 
our numerical procedure is rather economical 
we think that it is worthwhile
to pursue its application, e.g., to alternative versions
of the model or to similar self-consistency problems.

\section*{Acknowledgements}

H.S. thanks the Graduiertenkolleg "Er\-zeug\-ung und Zerf\"alle von 
Elementar\-teil\-chen" for financial support.

\begin{appendix}
\section{Partial waves in the $K$ spin basis}\label{anhangXi}
\setcounter{equation}{0}
The expansion with respect to K-spin harmonics $\Xi_{ij}$
\cite{Kahana:1984}
 \bea\label{greenexp}
\psi_{_{K,K_z,P}}(\bx)
=
\left(\begin{array}{rlcrl}
     u^{_{K,P}}_1(r)\!\!&\!\!\Xi^{_{K,K_z}}_{1}(\hbx)\!\!&\!\!+\!\!&\!\!
u_4^{_{K,P}}(r)\!\!
      &\!\!\Xi^{_{K,K_z}}_{4}(\hbx)\\
     u_2^{_{K,P}}(r)\!\!&\!\!\Xi^{_{K,K_z}}_{2}(\hbx)\!\!&\!\!+\!\!&\!\!
u_3^{_{K,P}}(r)\!\!
      &\!\!\Xi^{_{K,K_z}}_{3}(\hbx)
     \end{array}\right)\,\nonumber\\ 
\eea
written here for parity $(-1)^{(K+1)}$,
reduces the Dirac equation to radial equations for four coupled partial 
waves $u_i$.
The Hamiltonian acting on the radial wave functions
with parity $(-1)^{(K+1)}$
 via 
\be
  H^{K^P}_{ij}u_j=E u_i
\ee
is given by
\bea \nonumber
  {\cal{H}}^{K^P}&=&\left\{
  \begin{array}{cccc}
0&-\displaystyle\frac{d}{dr}-\frac{K+1}{r}&0&0\\
\displaystyle\frac{d}{dr}-\frac{K-1}{r}&0&0&0\\0&0&0&-
\displaystyle\frac{d}{dr}-\frac{K+2}{r}\\0&0&
\displaystyle\frac{d}{dr}-\frac{K}{r}&0\\
  \end{array}\right\}\\
&+&\left\{
  \begin{array}{cccc}
C(r)&-cS&sS(r)&0\\-cS(r)&-C(r)
&0&-sS(r)\\sS(r)&0&-C(r)&-cS(r)\\0&-sS(r)&-cS(r)&C(r)\\
  \end{array}\right\}\,,
\eea
where
\bea
S(r)&=&M\sin(\vartheta(r))\,,\\
C(r)&=&M\cos(\vartheta(r))\,,\\
s&=&2\sqrt{K(K+1)}/(2K+1)\,,\\
c&=&1/(2K+1)\,.
\eea
We ``square'' the Dirac equation 
to obtain an effective Klein-Gordon equation
\be
 \label{boswaveq}
  \left(-\Delta_{ij}^K+M^2\delta_{ij}+{\cal{V}}^K_{ij}\right)f_j=E^2 f_i\,,
\ee
where
\be
  \Delta^K_{ij}=\delta_{ij}\frac{1}{r^2}\frac{d}{dr}r^2\frac{d}{dr}-
\frac{K_i(K_i+1)}{r^2}\,.
\ee
The orbital angular momenta $K_i$ are given by
 $K_1=K-1$, $K_2=K_3=K$ and $K_4=K+1$.
The bosonized wave functions refer to the same K-spin basis
(\ref{greenexp}) as the fermionic ones.
The potential for the parity $(-1)^{K+1}$ is given by
\bea 
  \label{pottot}
&&{\cal{V}}^K=\\&&\nonumber
\left\{
  \begin{array}{cccc}\displaystyle
c\left(S'+\frac{2KS}{r}\right)&C'&0&\displaystyle
s\left(S'-\frac{S}{r}\right)\\
C'&c\displaystyle\left(-S'+\frac{2KS}{r}\right)&\displaystyle
s\left(S'+\frac{S}{r}\right)&0\\0&
\displaystyle s\left(S'+\frac{S}{r}\right)&
\displaystyle c\left(S'+\frac{2(K+1)S}{r}\right)&-C'\\
\displaystyle s\left(S'-\frac{S}{r}\right)&0&-C'&
\displaystyle c\left(-S'+\frac{2(K+1)S}{r}\right)\\
  \end{array}\right\}
\,,
\eea
where 
\bea
S'(r)&=&M\frac{d}{dr}\sin(\vartheta(r))\,,\\
C'(r)&=&M\frac{d}{dr}\cos(\vartheta(r))\,.
\eea
For the parity $(-1)^K$ the sign of the mass has to be changed.

While above we have formulated the Dirac and effective Klein-Gordon
equations for wave functions
with real minkowskian energies $E$ we mainly work
with euclidean mode functions whose argument we denote with
$\nu=-iE$. So the mode equations are analogous to
(\ref{boswaveq}) with $E^2 \to - \nu^2$. In term of these
mode finctions the bosonic Green function is given
by (\ref{gtheta}).
The fermionic euclidean mode functions are related to
the bosonized ones  via
\bea \label{modmod}
  u_1&=&\left(i\nu+C\right)f_{1}-\frac{K+1}{r}f_{2}-f_{2}'
                      -cSf_{2}+sSf_{3}\,,\\
  u_2&=&\left(i\nu-C\right)f_{2}-\frac{K-1}{r}f_{1}+f_{1}'
                      -cSf_{1}-sSf_{4}\,,\\
  u_3&=&\left(i\nu-C\right)f_{3}-\frac{K+2}{r}f_{4}-f_{4}'
                      -cSf_{4}+sSf_{1}\,,\\
  u_4&=&\left(i\nu+C\right)f_{4}-\frac{K}{r}f_{3}+f_{3}'
                      -cSf_{3}-sSf_{2}
\eea
for the parity $(-1)^{(K+1)}$, thus the Green function becomes complex.
With these radial functions the fermionic Green function
$S_E({\bf x},{\bf x}',\nu)$ 
in the K-Spin basis reads
\be\label{gthetatilde}
s_{nm}(r,r',\nu)=\kappa\left[
            \theta(r-r'){u}^{\alpha +}_n (\nu,r)
f^{\alpha -}_{m}(\nu,r') 
           +\theta(r'-r){u}^{\alpha -}_n (\nu,r)
f^{\alpha +}_{m}(\nu,r') 
            \right]\,.
\ee
\section{Some relations for K-spin harmonics}
\setcounter{equation}{0}
The action of $\tau_3$, 
$\sigma_3$ and $\sigma_3\tau_3$ on the K-spin harmonics
  \cite{Kahana:1984} is
\bea
 \tau_3\Xi^{K,K_z}_{1}&=&\frac{K_z}{K}\Xi^{K,K_z}_{1}-
\frac{\sqrt{K^2-K_z^2}}{K}\Xi^{K-1,K_z}_{3     }\,,\\
  \tau_3\Xi^{K,K_z}_{2}&=&\frac{K_z}{K}\Xi^{K,K_z}_{2}-
\frac{\sqrt{K^2-K_z^2}}{K}\Xi^{K-1,K_z}_{4     }\,,\\
  \tau_3\Xi^{K,K_z}_{3}&=&-\frac{K_z}{K+1}\Xi^{K,K_z}_{3}-
\frac{\sqrt{(K+1)^2-K_z^2}}{K+1}\Xi^{K+1,K_z}_{1     }\,,\\
  \tau_3\Xi^{K,K_z}_{4}&=&-\frac{K_z}{K+1}\Xi^{K,K_z}_{4}-
\frac{\sqrt{(K+1)^2-K_z^2}}{K+1}\Xi^{K+1,K_z}_{2      }\,,
\\ 
 \sigma_3\Xi^{K,K_z}_{1}&=
 &\frac{K_z}{K}\Xi^{K,K_z}_{1}
 -2\frac{\sqrt{K-1}\sqrt{K^2-K_z^2}}{(2K-1)\sqrt{K}}\Xi^{K-1,K_z}_{2}
 +\frac{\sqrt{K^2-K_z^2}}{(2K-1){K}}\Xi^{K-1,K_z}_{3}
 \,,\\ 
 \sigma_3\Xi^{K,K_z}_{2}&=&
 -2\frac{\sqrt{K}\sqrt{(K+1)^2-K_z^2}}{(2K+1)\sqrt{K+1}}\Xi^{K+1,K_z}_{1}
 -\frac{K_z(2K-1)}{K(2K+1)}\Xi^{K,K_z}_{2}
\\ \nonumber &&
 +2\frac{K_z}{(2K+1)\sqrt{K(K+1)}}\Xi^{K,K_z}_{3}
 -\frac{\sqrt{K^2-K_z^2}}{K(2K+1)}\Xi^{K-1,K_z}_{4}
 \,,\\
 \sigma_3\Xi^{K,K_z}_{3}&=
 &\frac{\sqrt{(K+1)^2-K_z^2}}{(2K+1)(K+1)}\Xi^{K+1,K_z}_{1}
  +2\frac{K_z}{(2K+1)\sqrt{K(K+1)}}\Xi^{K,K_z}_{2}
 \\ \nonumber&&
 +\frac{K_z(2K+3)}{(2K+1)(K+1)}\Xi^{K,K_z}_{3}
 -2\frac{\sqrt{K^2-K_z^2}\sqrt{K+1}}{(2K+1)\sqrt{K}}\Xi^{K-1,K_z}_{4}
 \,,\\
 \sigma_3\Xi^{K,K_z}_{4}&=
 &-\frac{\sqrt{(K+1)^2-K_z^2}}{(2K+3)(K+1)}\Xi^{K+1,K_z}_{2}
 \\ \nonumber&&
 -2\frac{\sqrt{(K+1)^2-K_z^2}\sqrt{K+2}}{(2K+3)\sqrt{K+1}}\Xi^{K+1,K_z}_{3}
 -\frac{K_z}{K+1}\Xi^{K,K_z}_{4}
 \,,\\
 \sigma_3\tau_3\Xi^{K,K_z}_{1}&=
 &-\frac{K-2K_z^2}{K(2K-1)}\Xi^{K,K_z}_{1}
 -2\frac{K_z\sqrt{K^2-K_z^2}}{(2K-1)\sqrt{K(K-1)}}\Xi^{K-1,K_z}_{2}
\\  \nonumber&&
 -2\frac{K_z\sqrt{K^2-K_z^2}}{K(2K-1)}\Xi^{K-1,K_z}_{3}
 +2\frac{\sqrt{(K^2-K_z^2)((K-1)^2-K_z^2)}}{(2K-1)\sqrt{K(K-1)}}
\Xi^{K-2,K_z}_{4}
 \,,\\ 
 \sigma_3\tau_3\Xi^{K,K_z}_{2}&=
 &-2\frac{K_z\sqrt{(K+1)^2-K_z^2}}{(2K+1)\sqrt{K(K+1)}}\Xi^{K+1,K_z}_{1}
 +\frac{K-2K_z^2}{K(2K+1)}\Xi^{K,K_z}_{2}
\\ \nonumber&&
 +2\frac{K^2+K-K_z^2}{(2K+1)\sqrt{K(K+1)}}\Xi^{K,K_z}_{3}
 +2\frac{K_z\sqrt{K^2-K_z^2}}{K(2K+1)}\Xi^{K-1,K_z}_{4}
 \,,\\ 
 \sigma_3\tau_3\Xi^{K,K_z}_{3}&=
 &-2\frac{K_z\sqrt{(K+1)^2-K_z^2}}{(2K+1)(K+1)}\Xi^{K+1,K_z}_{1}
 +2\frac{K^2+K-K_z^2}{(2K+1)\sqrt{K(K+1)}}\Xi^{K,K_z}_{2}
 \\ \nonumber&&
 -\frac{K+1+2K_z^2}{(2K+1)(K+1)}\Xi^{K,K_z}_{3}
 +2\frac{K_z\sqrt{K^2-K_z^2}}{(2K+1)\sqrt{K(K+1)}}\Xi^{K-1,K_z}_{4}
 \nonumber\,,\\
 \sigma_3\tau_3\Xi^{K,K_z}_{4}&=
 & 2\frac{\sqrt{((K+1)^2-K_z^2)
 ((K+2)^2-K_z^2)}}{(2K+3)\sqrt{(K+1)(K+2)}}\Xi^{K+2,K_z}_{1}
 \\&&
 +2\frac{K_z\sqrt{(K+1)^2-K_z^2}}{(2K+3)(K+1)}\Xi^{K+1,K_z}_{2}
 \nonumber\\&&
 +2\frac{K_z\sqrt{(K+1)^2-K_z^2}}{(2K+3)\sqrt{(K+1)(K+2)}}\Xi^{K+1,K_z}_{3}
 +\frac{1+K+2K_z^2}{(2K+3)(K+1)}\Xi^{K,K_z}_{4}
 \nonumber\,.
\eea
\end{appendix}

\newpage

\section*{ Figure captions}

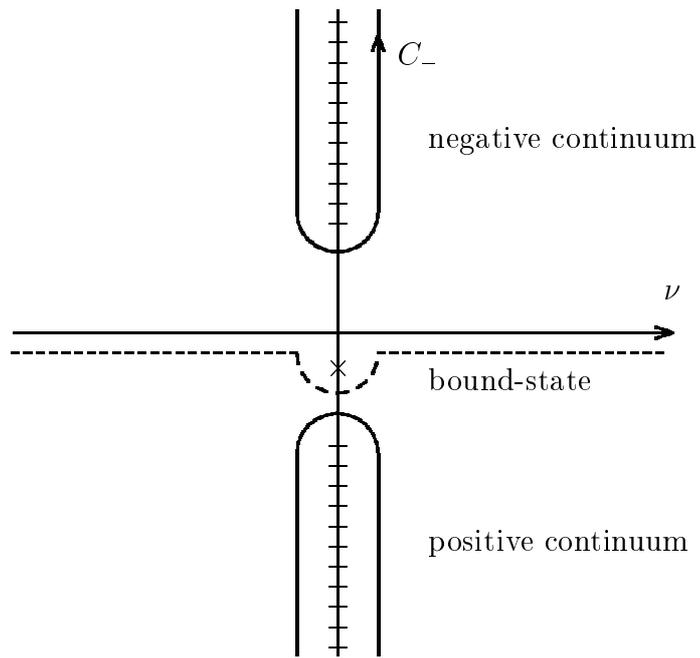
{\bf Fig. 1} : The complex $\nu$-plane: Solid line: the contour 
$C_-$ around the negative continuum states; dashed line: 
the deformed contour along the real $\nu$ axis, including, in addition, 
the bound state.
\\\\
{\bf Fig. 2a} : Asymptotic behavior of partial wave sums
and $\nu$ integrands: we display the behaviour of
the single contributions to the sum over partial waves 
for the $\nu$ integrand of the
energy at $\nu=1$. The continous line
is the power fit as described in the text.
\\\\
{\bf Fig. 2b} : the same as Fig. 2a, partial wave 
contributions to  the
$\nu$ integrand for $g_A$ at $\nu=0$.
\\\\
{\bf Fig. 2c} : the same as Fig. 2a, the $\nu$ integrand
for the energy.
\\\\
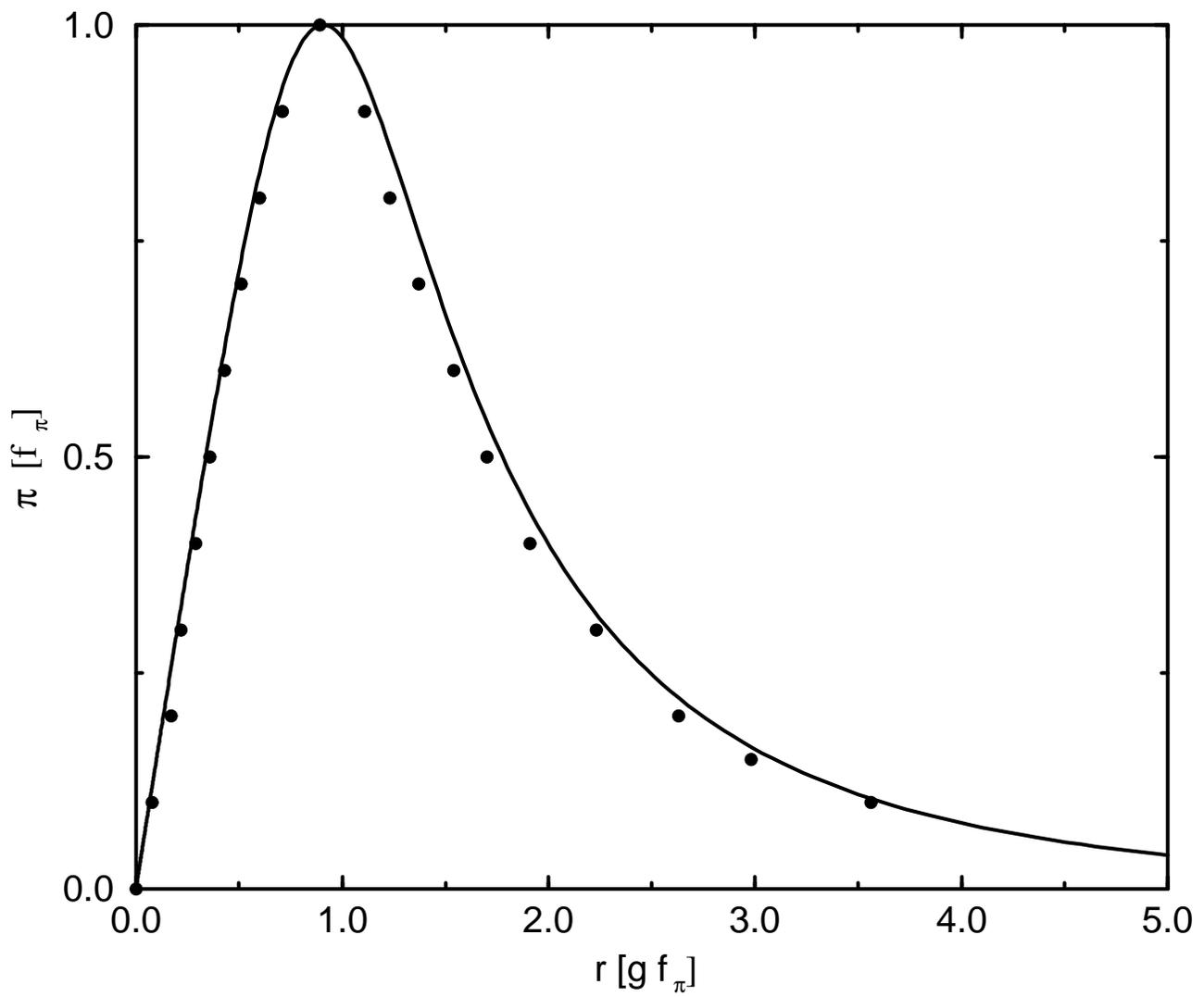
{\bf Fig. 3} : The mesonic profile of the nucleon for $g=4$.
We display the pion field $\pi=f_\pi\sin(\vartheta)$ as a 
function of $r$ for $g=4.0$ (solid line). For comparison
we also plot the result obtained previously 
by D\"oring {\em et al.} [5] using a different
technique (dots).
\\\\
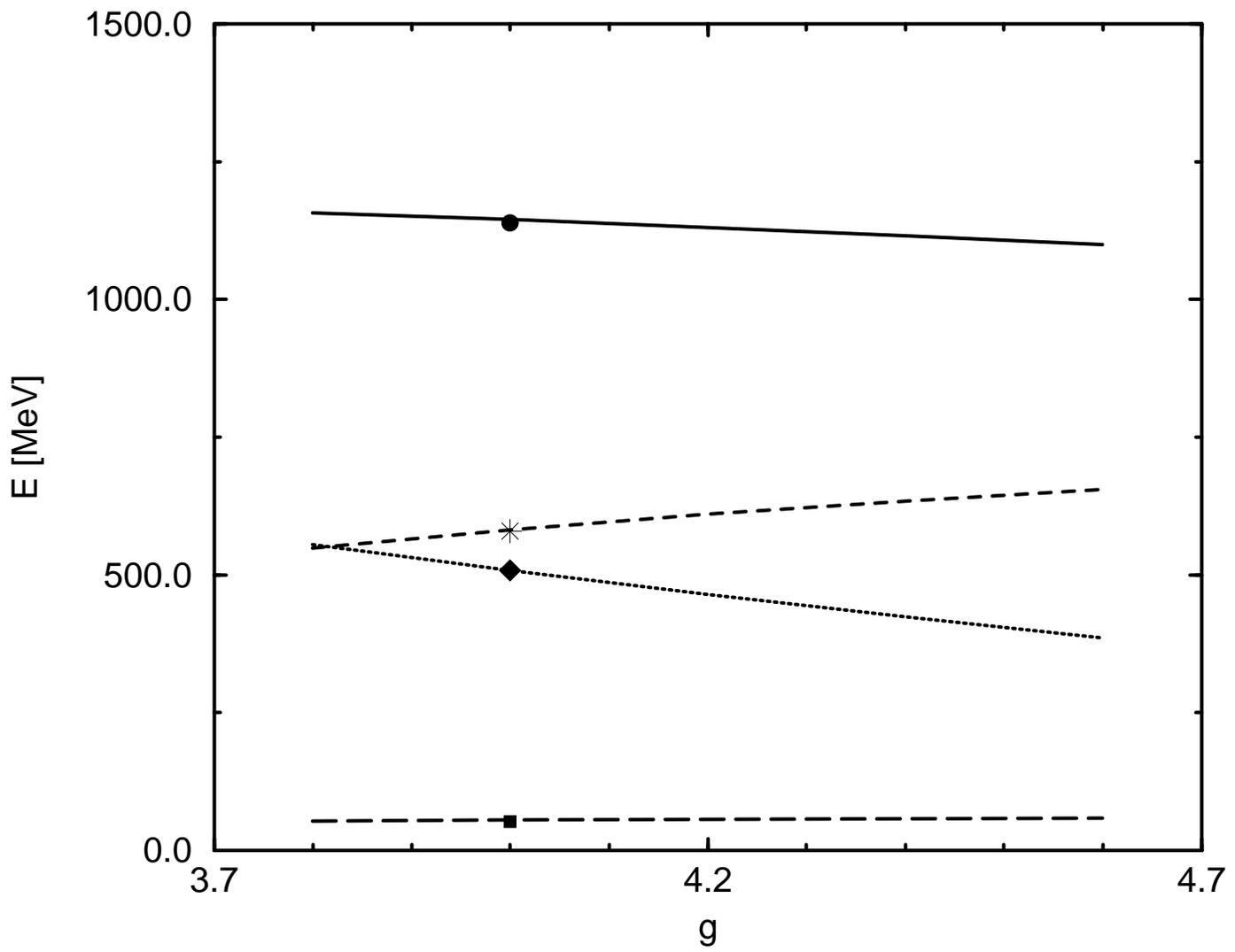
{\bf Fig. 4} : The energy of the nucleon.
We display the energy as a function of $g$ (solid line), 
the bound state energy (dotted line), the zero point energy (dashed line) and 
the energy of the symmetry breaking term (long dashed line).
For comparison we also plot the total energy and its parts 
obtained previously by D\"oring {\em et al.} [5] 
for $g=4$ (dot, square, diamond and cross).
\\\\
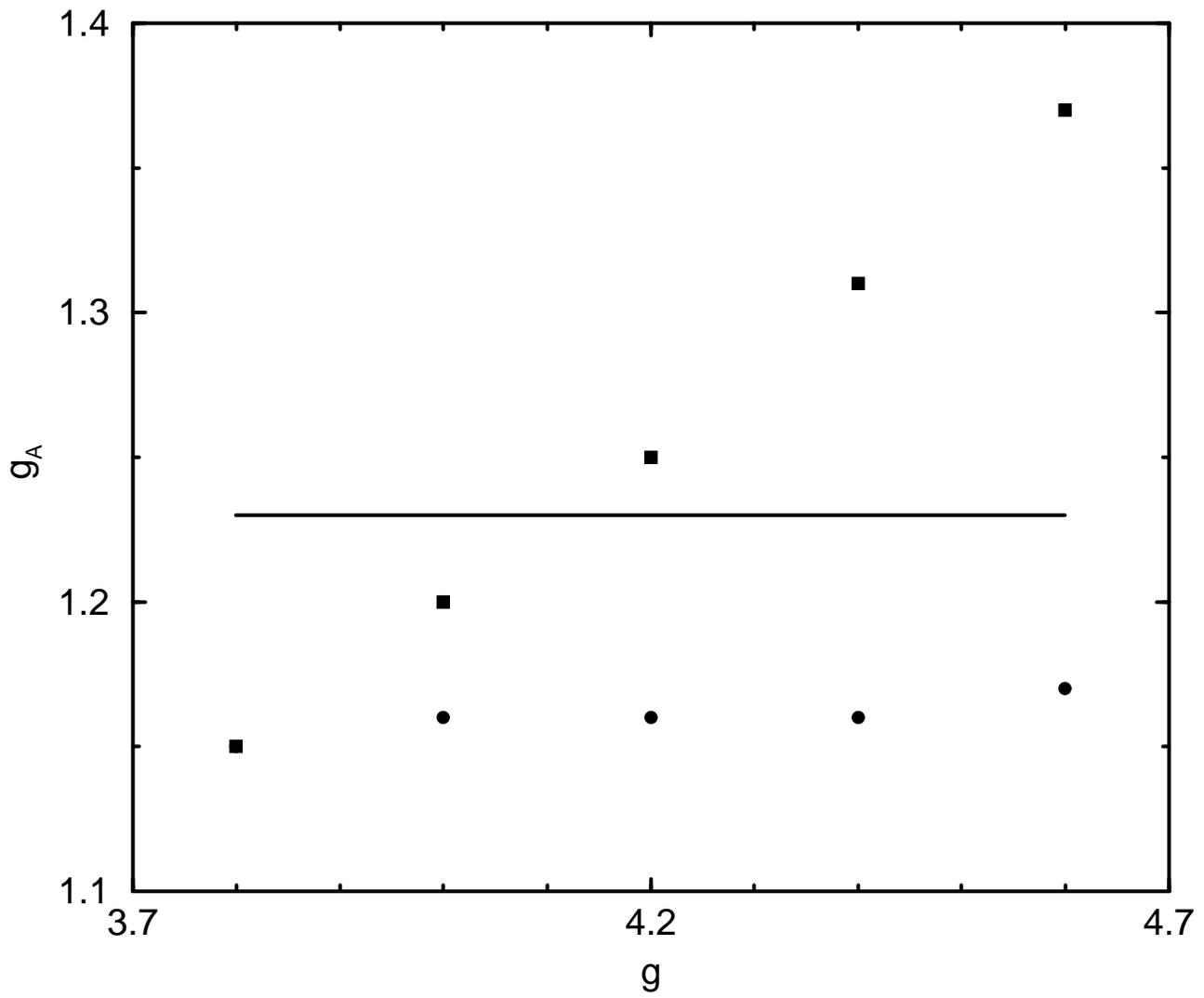
{\bf Fig. 5} : The axial vector coupling constant.
We display the  axial vector coupling constant
as a function of $g$. The squares are obtained by calculating the expectation 
value, the dots by using the Goldberger-Treiman relation.
The solid line is the experimental value.
\\\\
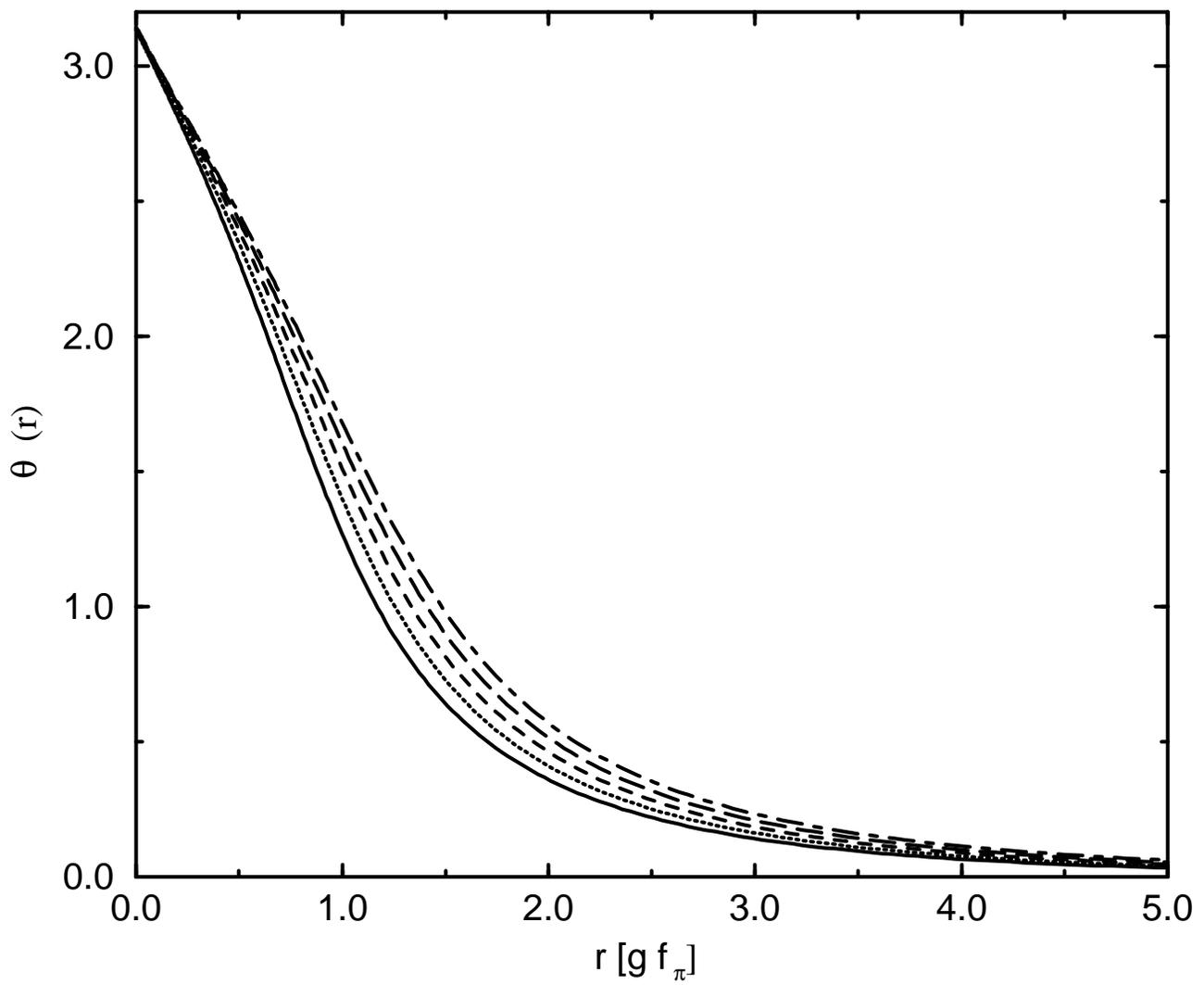
{\bf Fig. 6} : The mesonic profile for various couplings.
We display the mesonic profile $\vartheta$ as a 
function of $r$ for different coupling constants.
($g=3.8$ solid line, $g=4.0$ dotted line, $g=4.2$ dashed line,
$g=4.4$ long dashed line, $g=4.6$ dash-dotted line).
\\\\
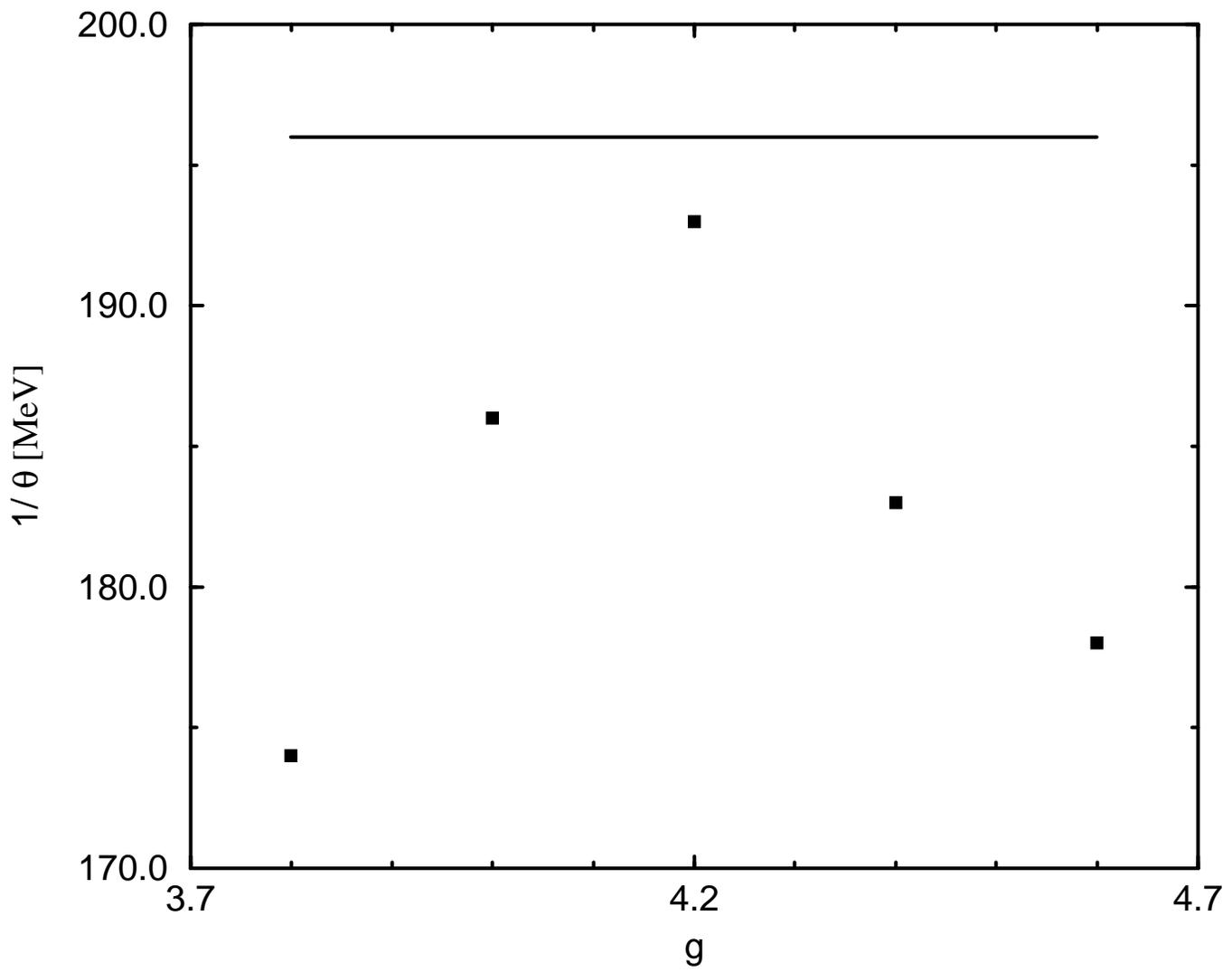
{\bf Fig. 7} : The moment of inertia.
We display the moment of inertia $\theta$ as a 
function for different coupling constants (dots).
The solid line is the experimental value.

\newpage

\begin{table}[htbp]
  \begin{center}
    \leavevmode
    \begin{tabular}{|lc|r|r|r|r|r|r|r|}
      \hline
      $g$&&3.8&4.0&4.2&4.4&4.6&4.0 \cite{Doring:1992sj}&Exp.\\
      \hline
      $E\sub{bou}$&[MeV]&555&508&464&424&386&507&\\
      $E\sub{con}$&[MeV]&549&582&610&634&655&579&\\
      $\Sigma$&[MeV]&53&55&57&58&58&52&45$\pm$9
      \cite{Gasser:1991ce}\\
      $E\sub{total}$&[MeV]&1157&1145&1131&1116&1099&1139&\\
      $1/\theta\top{b-c}$&[MeV]&199&218&233&247&259&&\\
      $1/\theta\top{c-c}$&[MeV]&1404&1264&1141&711&575&&\\
      $1/\theta\sub{total}$&[MeV]&174&186&193&183&178&&\\
      $M_\Delta-M\sub{N}$&[MeV]&261&279&290&274&267&&294\\
      $M\sub{N}$&[MeV]&1222&1214&1203&1184&1166&&939\\
      $\langle R^2\rangle\sub{bou}$&[fm]&0.56&0.56&0.58&0.60&0.62&&\\
      $\langle R^2\rangle\sub{con}$&[fm]&0.001&0.002&0.003&0.004&0.005&&\\
      $\langle R^2\rangle\sub{total}$&[fm]&0.56&0.56&0.58&0.60&0.62&0.58&0.62\\
      $g\sub{A}\top{bou}$&&0.72&0.71&0.71&0.70&0.70&0.71&\\
      $g\sub{A}\top{con}$&&0.43&0.49&0.55&0.61&0.67&&\\
      $g\sub{A}\top{total}$&&1.15&1.20&1.25&1.31&1.37&&1.23\\
      $g\sub{\pi NN}$&&11.6&11.7&11.8&11.8&11.8&&13.1 \cite{Matsinos:1998wp}\\
      $g\sub{A}\top{G-T}$&&1.15&1.16&1.16&1.16&1.17&0.96&1.23\\
      \hline
    \end{tabular}
  \end{center}
\caption{Static properties of the nucleon}
\label{tab1}
\end{table}
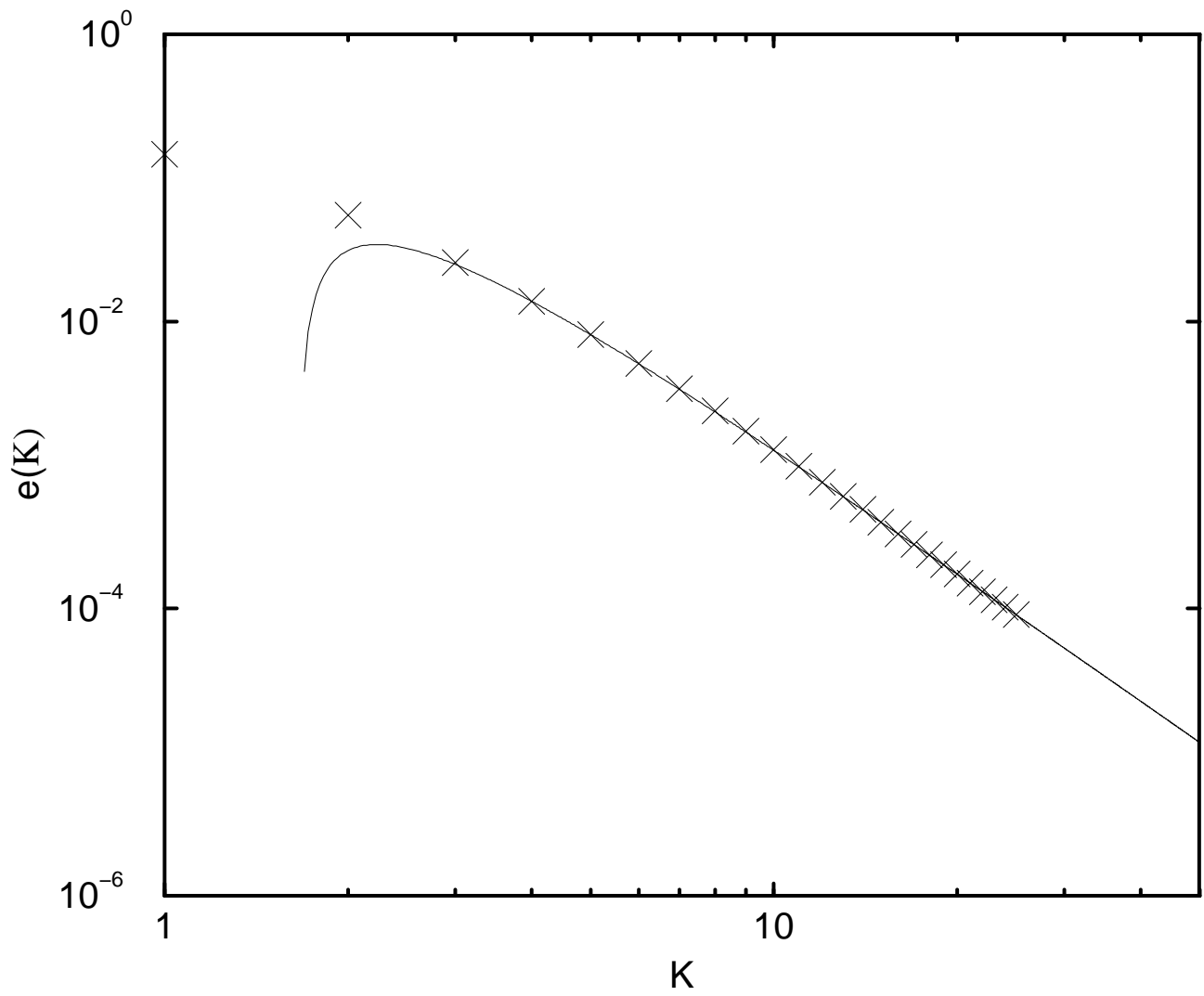
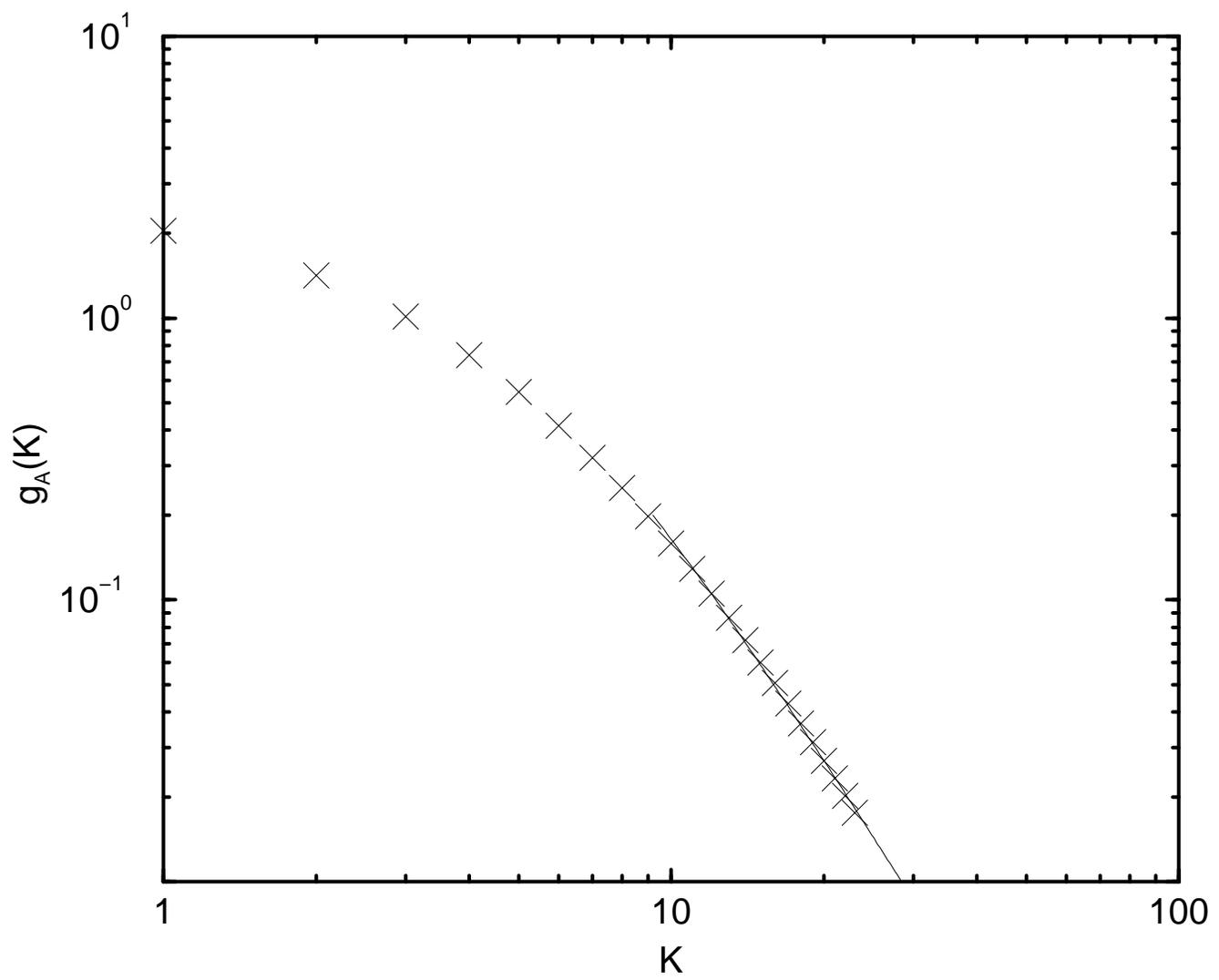
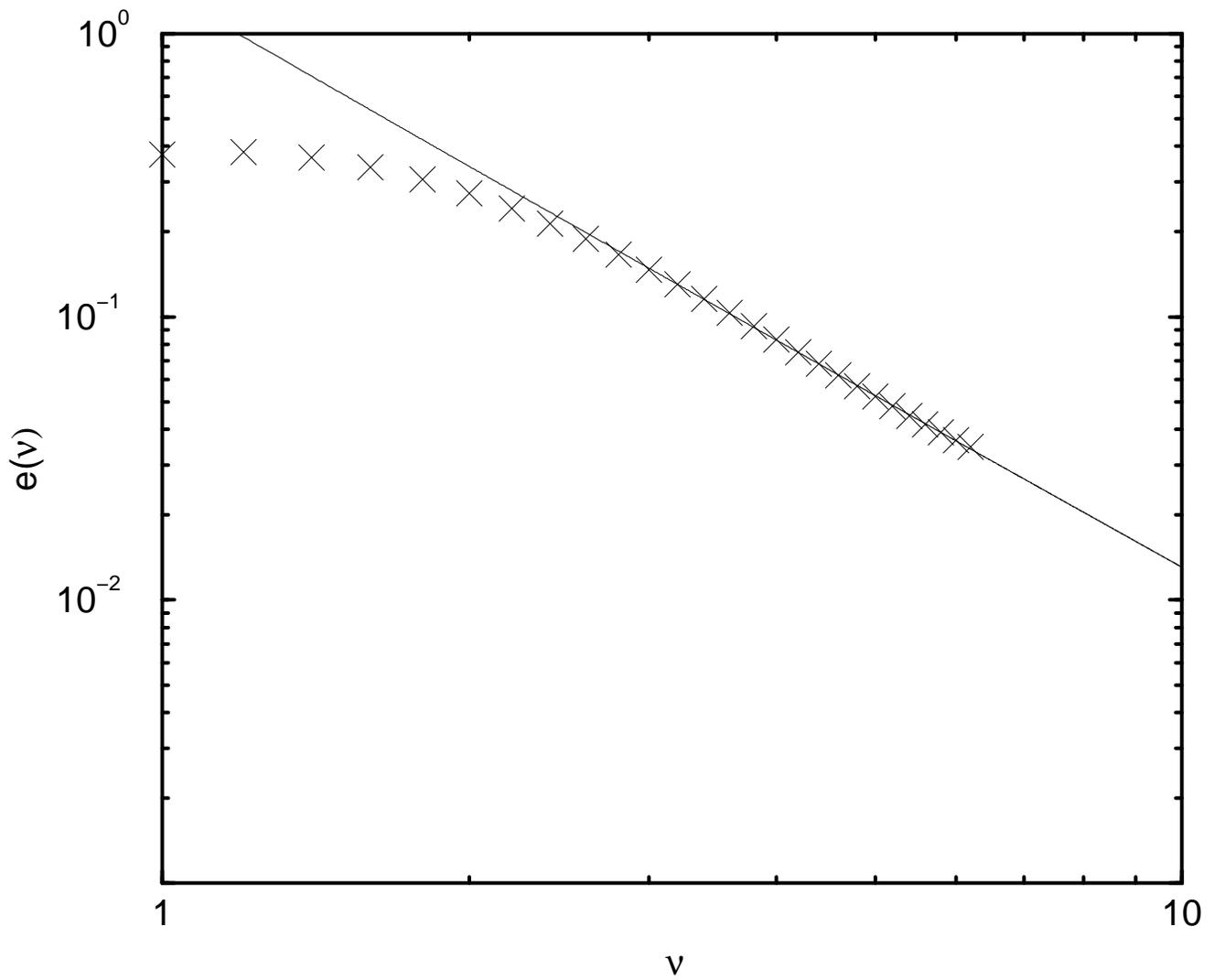
\end{document}